\documentclass[aps,twocolumn,showpacs,floatfix,groupedaddress,amsmath,amssymb,prx]{revtex4-1}
\usepackage{graphicx}
\usepackage{multirow}

\newcommand{\ket} [1] {| #1 \rangle}
\newcommand{\bra} [1] {\langle #1 |}

\newcommand{\ketbra}[2]{\ket{#1}\bra{#2}}

\newcommand{\mypsi}{\ket{\Psi^{\tiny \mbox{bound}}}}
\newcommand{\myphi}{\ket{\Psi^{\tiny \mbox{bulk}}}}

\newcommand{\myomega}{\ket{\Omega^{\tiny \mbox{bulk}}}}

\newcommand{\rhoy}{\hat{\rho}}

\newcommand{\openstate}{\ket{\Psi^{\tiny \mbox{open}}}}
\newcommand{\bondstate}{\ket{\Psi^{\tiny \mbox{bond}}}}

\newcommand{\xbound}[1]{{\langle#1\rangle}_{\tiny \mbox{bound}}}
\newcommand{\xbulk}[1]{{\langle#1\rangle}_{\tiny \mbox{bulk}}}

\begin{document}
\title{Tensor network state correspondence and holography}
\author{Sukhwinder Singh}
\affiliation{$^{1}$\mbox{Institute for Quantum Optics  \& Quantum Information, Austrian Academy of Sciences, Vienna, Austria}}
\email{\texttt{Sukhbinder.Singh@oeaw.ac.at}}
\affiliation{$^{2}$Center for Engineered Quantum Systems, Department of Physics \& Astronomy, Macquarie University, 2109 NSW, Australia}


\newcommand{\bulkPastCone}[1]{\mathcal{C}^{\tiny \mbox{bulk}}_{-}(#1)}
\newcommand{\bulkFutureCone}[1]{\mathcal{C}^{\tiny \mbox{bulk}}(#1)}

\newcommand{\eref}[1]{Eq.~(\ref{#1})}
\newcommand{\erefs}[2]{Eqs.~(\ref{#1})-(\ref{#2})}
\newcommand{\fref}[1]{Fig.~\ref{#1}}


\begin{abstract}
In recent years, tensor network states have emerged as a very useful conceptual and simulation framework to study quantum many-body systems at low energies. In this paper, we describe a particular way in which any given tensor network can be viewed as a representation of two different quantum many-body states. The two quantum many-body states are said to correspond to each other by means of the tensor network.We apply this ``tensor network state correspondence''---a correspondence between quantum many-body states mediated by tensor networks as we describe---to the multi-scale entanglement renormalization ansatz (MERA) representation of ground states of one dimensional (1D) quantum many-body systems. Since the MERA is a 2D hyperbolic tensor network (the extra dimension is identified as the length scale of the 1D system), the two quantum many-body states obtained from the MERA, via tensor network state correspondence, are seen to live in the bulk and on the boundary of a discrete hyperbolic geometry. The bulk state so obtained from a MERA exhibits interesting features, some of which caricature known features of the holographic correspondence of String theory. We show how (i) the bulk state admits a description in terms of ``holographic screens'', (ii) the conformal field theory data associated with a critical ground state can be obtained from the corresponding bulk state, in particular, how pointlike boundary operators are identified with
extended bulk operators. (iii) We also present numerical results to illustrate that bulk states, dual to ground states of several critical spin chains, have exponentially decaying correlations, and that the bulk correlation length generally decreases with increase in central charge for these spin chains.
\end{abstract}

\maketitle
\tableofcontents
\section{Introduction}\label{sec:intro}

Low energy states of local Hamiltonians often contain entanglement that is limited in a way that can be exploited to efficiently parameterize such states by means of a \textit{tensor network}. 
A tensor network representation of a quantum many-body state consists of a set of tensors (multi-dimensional arrays of complex numbers) that are interconnected according to a given geometry. The \textit{open indices}---indices that are attached to only tensor---of the tensor network are associated with the physical degrees of freedom of the system. On the other hand, the \textit{bond indices}---indices that connect the tensors in the network---are associated with auxiliary degrees of freedom that carry the entanglement and correlations in the quantum many-body state. The specific way in which the tensors are connected by the bond indices, that is, the geometry of the tensor network bounds the entanglement entropy of the state \cite{TNGeometry}. On the other hand, these auxiliary degrees of freedom are summed over in order to obtain expectation values of observables from the tensor network. 

In recent years, the formalism of tensor networks has led to a wealth of practical simulation and conceptual tools to study quantum many-body systems at low energies. Examples of popular tensor network states include matrix product states \cite{Fannes92} (MPS), projected entangled pair states \cite{Verstraete04} (PEPS), tree tensor networks \cite{Shi06,Tagliacozzo09} (TTN), and the multi-scale entanglement renormalization ansatz \cite{ER,MERA} (MERA). These tensor network states form the basis of several powerful algorithms for efficient simulation of quantum many-body states, see e.g. Refs.\onlinecite{White92,Shi06,MERA,MERAcrit,Verstraete04,Tagliacozzo09}. In this paper, we propose a new interpretation of the auxiliary degrees (associated with the bond indices) of freedom in a tensor network representation, motivated by recent connections between the MERA and the AdS/CFT correspondence, which we briefly review below. 
 
The MERA representation of the ground state of a $d$-dimensional quantum lattice system is a $d+1$-dimensional tensor network that encodes the renormalization group (RG) flow of the ground state (and the Hamiltonian) \cite{ER}. Specifically, the extra dimension of the tensor network corresponds to length scale of the $d$ dimensional system, in the sense that different parts of the tensor network capture properties of the ground state at different length scales. The MERA has been successfully applied to simulate ground states of several quantum lattice systems, and seems particularly well suited for critical systems in $d=1$ dimensional systems. Critical systems are described by conformal field theories (CFTs) in the continuum, and the properties of the CFT such as its scaling dimensions, operator product expansion coefficients etc. can be easily obtained from the MERA representation of the CFT ground state \cite{MERAcrit}. 

On the other hand, the AdS/CFT correspondence \cite{AdSCFT,AdSCFTdictionary} is an equivalence between certain quantum gravity theories in $d+1$-dimensional AdS spacetimes and CFTs that live on the $d$-dimensional boundary of the AdS spacetime. The correspondence essentially translates the RG description of the CFT to a bulk gravity theory in AdS geometry. In particular, the extra dimension of the AdS geometry is identified with length scale in the CFT. The most studied examples of the AdS/CFT correspondence consist of CFTs with a large central charge which are dual to classical gravity in the bulk. In these examples, the central charge of the CFT sets the radius of curvature of the AdS geometry.  On the other hand, a small boundary central charge  (e.g. of order 1) often corresponds to \textit{quantum} gravity in the bulk \cite{quantumFlucBulk}. For example, Ref.~\onlinecite{IsingGravity} presents a holographic description of the 1D quantum critical Ising model, which has central charge equal to $\frac{1}{2}$. Indeed, there the authors match the partition function of the Ising model to a dual quantum gravity partition function, obtained by summing over all bulk geometries (gravitational fields) that are compatible with the asymptotic constraints imposed by the boundary theory.

The AdS/CFT correspondence consists of a concrete prescription of obtaining the boundary correlation functions from the bulk. More specifically, $n$-point correlators of the boundary are obtained evaluating Witten diagrams in the bulk, which describe certain scattering processes along extended trajectories in the bulk \cite{AdSCFTdictionary,WittenDiagrams}. 
The AdS/CFT correspondence also relates certain properties of the CFT to properties of the bulk gravity theory. For example, a CFT scaling field with scaling dimension $\Delta$ corresponds to a bulk field of mass $\Delta$, a global symmetry of the CFT generally translates to a local gauge symmetry in the bulk etc. Another example is the Ryu-Takayanagi formula, which holds when the bulk is described by classical gravity \cite{HoloEntropy}. For instance, for an $1+1$ dimensional CFT that has a classical gravity dual, the Ryu-Takayanagi formula equates (in appropriate units) the von Neumann entanglement entropy of an intervalin the CFT vacuum to the length of the geodesic that extends between the end points of the interval through the dual bulk spacetime. (An analogous holographic interpretation of Reyni entanglement entropy has also been proposed \cite{HoloEntropy1}.) 

The basic premise of the MERA-AdS/CFT conjecture is that the MERA tensor network has a hyperbolic geometry---which may be interpreted as a spatial slice of an AdS geometry---and the entanglement entropy of an interval is accounted by means by bond indices that are intersected by geodesics in the hyperbolic geometry, which appears to be qualtitatively similar to the Ryu-Takayanagi formula \cite{Swingle}. Since this first observation, the conjecture has been elaborated by several authors \cite{MERAHolo,deSitter,localScaleMERA,ExactHolo,HoloCode}.
While it is understood how the MERA encodes an 1D critical ground state, there is no settled view on how it could also encode a dual description of an emergent 2D system, though some  interesting proposals have appeared recently \cite{ExactHolo,HoloCode,HoloRandom}. In particular, the identification of suitable bulk degrees of freedom remains an open question.

In this paper, we describe how \textit{any} tensor network (with open indices) may be viewed as a relation or a `correspondence' between quantum many-body states that live in different Hilbert spaces. This is achieved by treating the auxiliary degrees of freedom included the tensor network description as physical degrees of freedom of an extended system, which allows one to view the same tensor network as describing another quantum many-body state of a larger Hilbert space. (In other words, we view a tensor network as representing two different states by associating a many-body Hilbert space with the tensor network in two different ways.)  We advocate, and illustrate with concrete examples, the view that the two quantum many-body states, which are encoded in the same tensor network but in two different ways, must be systematically related together. We refer to this correspondence as \textit{tensor network state correspondence}.

We then apply this correspondence to the MERA to obtain two quantum many-body states, which are seen to live on the boundary and in the bulk of a manifold respectively. We illustrate how the 2D bulk states obtained from the MERA representation of 1D critical (CFT) ground states caricature certain features of the AdS/CFT correspondence \cite{strongadscft}.
For example, entanglement in a bulk state is organized according to `holographic screens', analogous to holographic horizons that appear in quantum gravity models. Here, a holographic screen consists of bulk sites that are located along a 1D path anchored at two boundary locations, such that the entanglement entropy of the sites located on the path is \textit{equal} to the entanglement entropy of the 2D bulk region enclosed between the path and the boundary of the geometry. We show that the $n$-point correlators of a critical ground state represented by the MERA can be obtained from the expectation value of \textit{extended} operators in the dual bulk state. This caricatures the prescription to calculate boundary correlators in AdS/CFT by evaluating Witten diagrams.

Here we do not attempt to deduce a semi-classical bulk geometry from the dual bulk states obtained from the MERA, or assume any particular spacetime interpretation of the MERA's hyperbolic geometry. Instead, we focus on the entanglement and correlations properties of the bulk states. This is particularly relevant for holographic duals of critical Hamiltonians that described by a CFT with a small central charge, where the bulk does not admit a description in terms of a classical spacetime. (In fact, the MERA has been mostly applied to simulate ground states of CFTs that have a small central charge.) 
On the other hand, bulk states may encode classical geometry in the limit in large central charge. However, we will not explore this important issue here.

Finally, we remark that while we will draw upon on certain structural features of the AdS/CFT correspondence in this paper, the discussion is more focused on tensor network aspects. Our paper is best followed as a new bulk/boundary correspondence that is derived from tensor networks, but which bears several non-trivial resemblances to the AdS/CFT correspondence.

\subsection{Summary of the main results}
The paper is organized as follows. 


In Sec.~\ref{sec:tncgeneral}, we introduce tensor network state correspondence for a generic tensor network representation, namely, how any given tensor network representation can be viewed as a correspondence or relation between two quantum many-body states. The two states are obtained from by associating two different many-body Hilbert spaces to the Hilbert space. In our view, this is quite a natural dual description of any tensor network representation, since it is essential based on treating the bond indices on the same footing as the open indices. 

After introducing the correspondence for a generic tensor network, we turn to addressing the main goal of the paper. In Sec.~\ref{sec:boundary}, we briefly review the MERA representation of ground states of infinite 1D quantum lattice systems.  In Sec.~\ref{sec:bulk}, we apply the tensor network state correspondence to the MERA to obtain two states that can be viewed as living in the bulk and on the boundary of a manifold respectively. In light of the on-going dialogue between the MERA and holography, we propose that the bulk state is the `holographic dual' of the boundary state. To this end, we describe some interesting features of certain bulk states, which appear to caricature features of the holographic correspondence. These are:

(i) The entanglement in MERA bulk states is organized according to holographic screens. Namely, expectation values and entanglement entropy of certain 2D regions in the bulk can be obtained by considering only the 1D boundary of the region. This appears to caricature the presence of holographic horizons in quantum gravity models. (Sec.~\ref{sec:bulkholo}.)

(ii) The correlation functions of a critical ground state, corresponding to the vacuum of a CFT in the continuum, can be obtained from the expectation values of certain extended operators in certain the dual bulk state. This appears to caricature the prescription in AdS/CFT to obtain correlators of the CFT by evaluating Witten diagrams in the bulk. (Sec.~\ref{sec:dictionary}.)

(iii) For a boundary critical ground state, we provide numerical evidence to show that the bulk entanglement entropy is smaller for larger central charge. (The central charge is a property of the CFT that describes the critical ground state in the vacuum.) This is compatible with the fact that in AdS/CFT, the leading order of quantum fluctuations in the bulk is inverse central charge. (Sec.~\ref{sec:bulkentanglement}.)

(iv) We also prove that the dual bulk states, derived from the MERA, exhibit an area law entanglement scaling (see Appendix \ref{app:arealaw}). This scaling is found commonly in ground states of quantum many-body systems whose dynamics is governed by local Hamiltonians.

In a following paper, Ref.~\onlinecite{TNCSym}, we generalize this correspondence to accomodate the presence of a global onsite symmetry at the boundary. In particular, we describe how the global boundary symmetry gets gauged in the bulk, namely, the boundary state has a global symmetry while the dual bulk state has a local gauge symmetry. The bulk gauging of global boundary symmetries appears as a general rule of thumb in the AdS/CFT dictionary.

We conclude this section by remarking that our holographic dual description of the MERA differs from the Qi's proposal dubbed \textit{Exact Holographic Mapping} \cite{ExactHolo}. In the latter, the dual bulk degrees of freedom are associated with the tensors of the MERA, by introducing open indices on the tensors, while in our construction the dual bulk degrees of freedom are associated with the bonds of the MERA. Our construction crucially differs from Qi's proposal, since e.g. it leads to bulk correlations that are organized according to holographic screens, and also to conveniently introduce gauge transformations in the bulk, in order to realize the holographic gauging of the global boundary symmetry as described in Ref.~\onlinecite{TNCSym}.

\section{A tensor network as a correspondence between two quantum many-body states}\label{sec:tncgeneral}

The basic idea behind tensor network state correspondence is quite simple, so we begin straightaway by describing it for a generic tensor network decomposition of a quantum many-body state before applying it to the MERA to construct a holographic correspondence in the remainder of the paper.

\begin{figure}
  \includegraphics[width=\columnwidth]{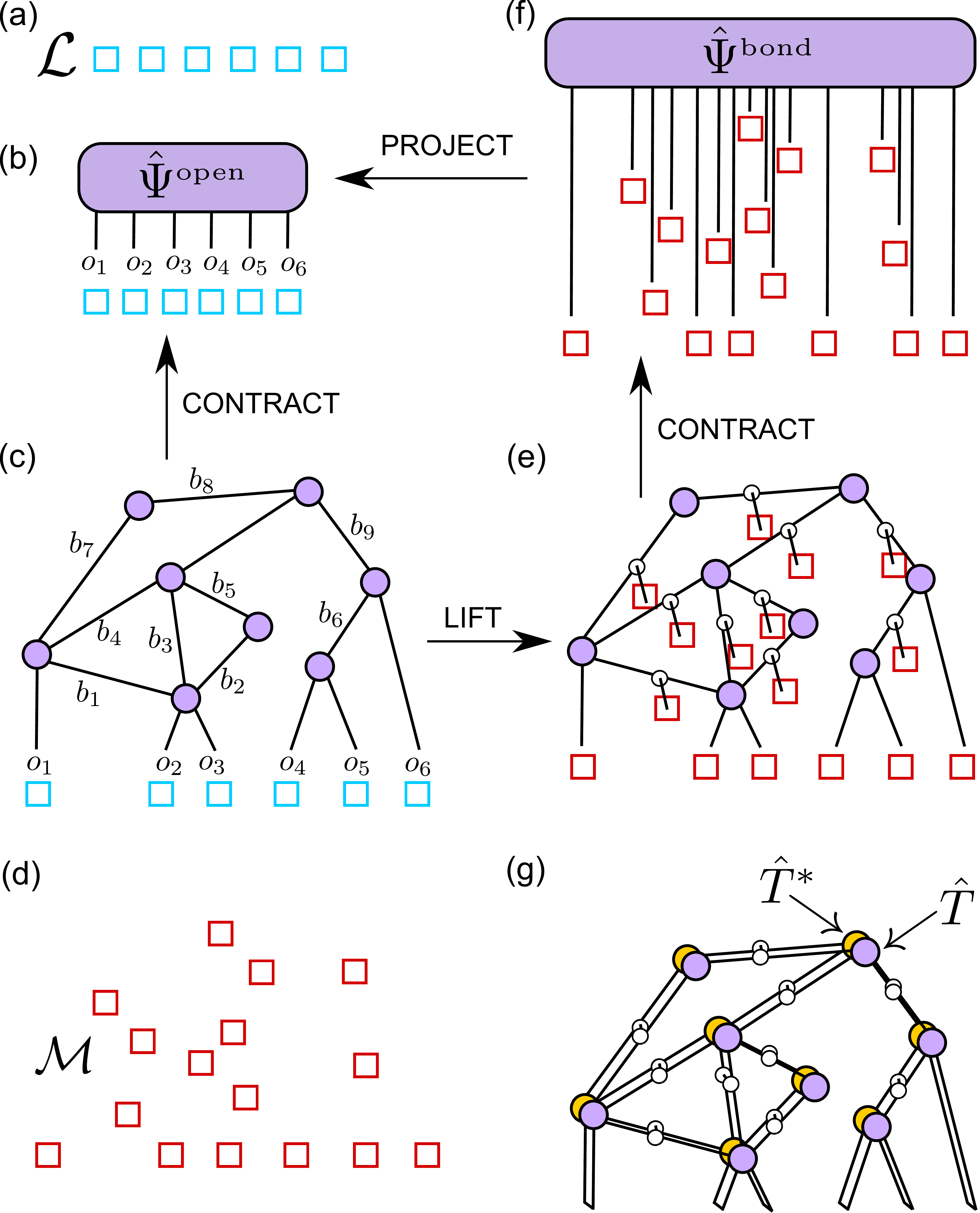}
\caption{\label{fig:tnduality} An illustration of tensor network state correspondence. (a). Lattice $\mathcal{L}$ with $L=6$ sites. (b) Tensor $\hat{\Psi}^{\mbox{\tiny open}}$ that encodes the probability amplitudes of a state $\openstate$ of the lattice $\mathcal{L}$, \eref{eq:open}, is represented by means of a tensor network $\mathcal{T}$. (c) Tensor network decomposition $\mathcal{T}$ of tensor $\hat{\Psi}^{\mbox{\tiny open}}$, where open indices $o_1,o_2,\cdots,o_6$ coincide with the indices of tensor $\hat{\Psi}^{\mbox{\tiny open}}$ and label a basis on the sites of $\mathcal{L}$. Tensor $\hat{\Psi}^{\mbox{\tiny open}}$ is recovered by contracting the tensor network $\mathcal{T}$, which involves summing over the bond indices $b_1,b_2,\cdots,b_9$. (d) An extended lattice $\mathcal{M}$ is composed by associating sites (shown in red) with all the indices of $\mathcal{T}$ (after embedding it in an ambient space), such that each index labels an orthornormal basis on the associated site. (e) The `lifted' tensor network obtained by inserting copy tensors on the bond indices of the tensor network $\mathcal{T}$. (f) Tensor $\hat{\Psi}^{\mbox{\tiny bond}}$ that is obtained by contracting all the tensors of the lifted tensor network, and whose components are the probability amplitudes of the bond state $\bondstate$, \eref{eq:bond}: a quantum state of the extended lattice $\mathcal{M}$. (g) Tensor network contraction that equates to the norm of the bond state $\bondstate$.}
\end{figure}

Consider a quantum state $\openstate$ of a lattice $\mathcal{L}$---made of $L$ sites, each described by a finite dimensional Hilbert space $\mathbb{V}$---given by
\begin{equation}\label{eq:genstate}
\openstate = \sum_{o_1,o_2,\ldots,o_L} \hat{\Psi}^{\tiny \mbox{open}}_{o_1o_2\ldots o_L}\ket{o_1}\otimes\ket{o_2}\otimes \ldots \ket{o_L},
\end{equation}
where $\{\ket{o_i}\}$ is an orthonormal basis on site $i$ and tensor $\hat{\Psi}^{\tiny \mbox{open}}$ decomposes into a tensor network $\mathcal{T}$, see \fref{fig:tnduality} (a)-(c). In this paper, by a tensor we simply mean a multi-dimensional array of complex numbers, and not necessarily an object with covariant and contravariant indices as used in relativity. A tensor network is a set of tensors that are interconnected according to a given (discrete) geometry (as described by a graph).

The superscript `open' in \eref{eq:genstate} indicates that the sites of lattice $\mathcal{L}$ are associated only with the \textit{open indices} of $\mathcal{T}$. An open index is attached to only one tensor in the network. The tensor network $\mathcal{T}$ has $L$ open indices $o_1,o_2,\cdots,o_L$, each of which labels an orthonormal basis on a different site of the lattice, \eref{eq:genstate}. The tensor network also has $N$ \textit{bond indices} $b_1,b_2,\cdots,b_N$, each of which connects a different pair of tensors. For simplicity, we assume that each bond index has the same size, namely, it ranges over the same number of values: $\{1,2,\cdots,\chi\}$.

For our purposes, we find it convenient to introduce the short-hand \textit{collective index} notation. For example, we re-write \eref{eq:genstate} as
\begin{equation}\label{eq:genstatecompact}
\openstate = \sum_o \hat{\Psi}^{\tiny \mbox{open}}_o \ket{o},
\end{equation}
where $o\equiv(o_1,o_2,\ldots,o_{L})$ denotes the \textit{collective open index}, namely, a tuple of values of all the open indices and $\ket{o}$ succintly denotes the tensor product basis $\ket{o_1}\otimes\ket{o_2}\otimes \cdots \otimes \ket{o_L}$. If each open index has size $d$ (that is, the dimension of the vector space $\mathbb{V}$ is $d$) then the size of the collective index $o$ is $d^L$.

Let $b\equiv(b_1,b_2,\ldots,b_{n_B})$ denote the \textit{collective bond index}. Then the state $\openstate$ can also be expanded as
\begin{equation}\label{eq:open}
\openstate = \sum_{o} (\sum_{b} \hat{\Phi}_{ob}) \ket{o},
\end{equation}
where $\hat{\Phi}_{ob}$ is a complex number obtained by multiplying together complex numbers from all tensors of $\mathcal{T}$ for given values of collective indices $o$ and $b$. The probability amplitudes $\hat{\Psi}^{\tiny \mbox{open}}_o$ in \eref{eq:genstatecompact} are related to the complex numbers $\hat{\Phi}_{ob}$ as
\begin{equation}\label{eq:tnstate}
\hat{\Psi}^{\tiny \mbox{open}}_{o} = \sum_b \hat{\Phi}_{ob}.
\end{equation}
State $\openstate$ that is encoded in a tensor network according to \eref{eq:tnstate} is called a \textit{tensor network state}. Recall that the MPS, TTN, PEPS and MERA are all examples of tensor network states.

Let us now extend the lattice $\mathcal{L}$ to a larger quantum many-body system $\mathcal{M}$ by introducing new sites, one for each bond index of the tensor network. Each bond site is described by a $\chi$-dimensional Hilbert space $\mathbb{W}$, and let $\{\ket{b_i}\}_{b_i=1}^\chi$ denote an orthornormal basis in $\mathbb{W}$. (This is the basis in which the components of the various tensors are expressed.) 

The same tensor network $\mathcal{T}$ can be regarded as a representation of a different quantum state $\bondstate$ of the extended quantum many-body system described by the Hilbert space  $\mathbb{W}^{(\otimes N)} \otimes \mathbb{V}^{(\otimes L)} (\equiv \mathcal{M})$ given by
\begin{equation}\label{eq:bond}
\bondstate = \sum_{o} \sum_{b} \hat{\Phi}_{ob} \ket{b} \otimes \ket{o},
\end{equation}
where $\hat{\Phi}_{ob}$ are the \textit{same} complex numbers that appear in \eref{eq:open} and $\ket{b} \equiv \ket{b_1} \otimes \ket{b_2} \otimes \cdots \otimes \ket{b_N}$. We refer to the state $\bondstate$ that is encoded in the tensor network $\mathcal{T}$ according to \eref{eq:bond} as a \textit{tensor network bond state}, or simply as a bond state.

A well known example of a bond state is a \textit{spin network state} that appears as a gauge-invariant basis state, for example, in lattice gauge theory \cite{BaezSpinNetwork} and loop quantum gravity \cite{LoopSpinNetwork}. A spin network state is represented by means of a \textit{spin network}---a tensor network composed of intertwiners of a compact Lie group and whose indices---\textit{both} open and bond---carry irreducible representations of the group and label a basis for a local degree of freedom of the system.

Another example of a bond state is a \textit{string-net state} \cite{StringNet} 
(a topological spin liquid) which is represented by means of a \textit{string-net}---a mathematical object very similar to a spin network, but where intertwiners and irreducible representations of a symmetry group are replaced with fusion vertex operators and charges of a Hopf algebra respectively. Once again, \textit{every} index of the string-net is associated with a local degree of freedom that collectively realize the spin liquid.

A bond state can also be regarded as a regular \textit{tensor network state} that is encoded in a tensor network obtained by modifying $\mathcal{T}$ by inserting a 3-index \textit{copy tensor} $\hat{c}: \mathbb{W} \rightarrow \mathbb{W} \otimes \mathbb{W}$ on each bond of the tensor network $\mathcal{T}$. In the bond basis (the basis chosen in the vector space $\mathbb{W}$) the copy tensor $\hat{c}$ is defined as
\begin{equation}\label{eq:copytensor}
\hat{c}_{pqr} = \delta_{pq}\delta_{qr} ~~~~~~~~~\mbox{(copy tensor)}.
\end{equation}
We refer to this modification as `lifting' the tensor network  $\mathcal{T}$, and the modified tensor network so obtained as the \textit{lifted} tensor network $\mathcal{T}$. By using each open index of the lifted tensor network to label an orthonormal basis on a different site of $\mathcal{M}$, the lifted tensor network can be regarded as representing a quantum state of $\mathcal{M}$. (Note that the open indices of the lifted indices correspond to both the open and bond indices of the original tensor network $\mathcal{T}$.) It is readily checked that the lifted tensor network represents the quantum state $\bondstate$ [\eref{eq:bond}], now according to \eref{eq:open}. More specifically, the amplitude for a given configuration $\ket{b} \otimes \ket{o}$ of the lattice $\mathcal{M}$ is obtained from the lifted tensor network by fixing the value of all the open indices corresponding to this configuration and contracting all the tensors; it is readily checked that the amplitude is equal to $\hat{\Phi}_{ob}$ [\eref{eq:bond}].

To summarize, we began with a quantum state $\openstate$ of a lattice $\mathcal{L}$, decomposed it into a tensor network according to \eref{eq:open}, and then constructed another quantum state $\bondstate$ from the same tensor network according to \eref{eq:bond} (this sequence is depicted by the trail of arrows in \fref{fig:tnduality}). Both states are encoded in the same tensor network, but in different ways.

The tensor network state $\openstate$ can be recovered from the tensor network bond state $\bondstate$ as
\begin{equation}\label{eq:recoverstate}
\openstate = (\bigotimes_{j=1}^{N} \hat{P}_{j}) \bondstate,~~~~~\hat{P}_{j} = \ketbra{+}{+},
\end{equation}
where $\hat{P}_{j}$ projects bond site $j$ of $\mathcal{M}$ to the state $\ket{+} = \frac{1}{\sqrt{\chi}}\sum_{j=1}^\chi \ket{j}$.

Thus, a given tensor network may be used to represent either a tensor network state, by using its open indices to label a basis of a quantum many-body system, or a tensor network bond state, by using both its open and bond indices to label a basis of an extended quantum many-body system. We will say that the states $\openstate$ and $\bondstate$ \textit{correspond} to each other by means of the tensor network $\mathcal{T}$, and denote it symbolically as
\begin{equation}\label{eq:tnsc}
\openstate \longleftarrow \mathcal{T} \longrightarrow \bondstate.
\end{equation}
We refer to this correspondence between the two quantum states $\openstate$ and  $\bondstate$ as \textit{tensor network state correspondence}. In the remainder of the paper, our goal is to describe how various properties of the two states $\openstate$ and  $\bondstate$ may be related to each other in the context of a specific tensor network, the MERA.

\section{A bulk/boundary correspondence from the MERA}\label{sec:bbcorrespondence}
In this section, we describe how tensor network state correspondence, when applied to the MERA, leads to a bulk/boundary type correspondence between two quantum many-body states.

\subsection{Boundary state}\label{sec:boundary}
\begin{figure}[t]
  \includegraphics[width=\columnwidth]{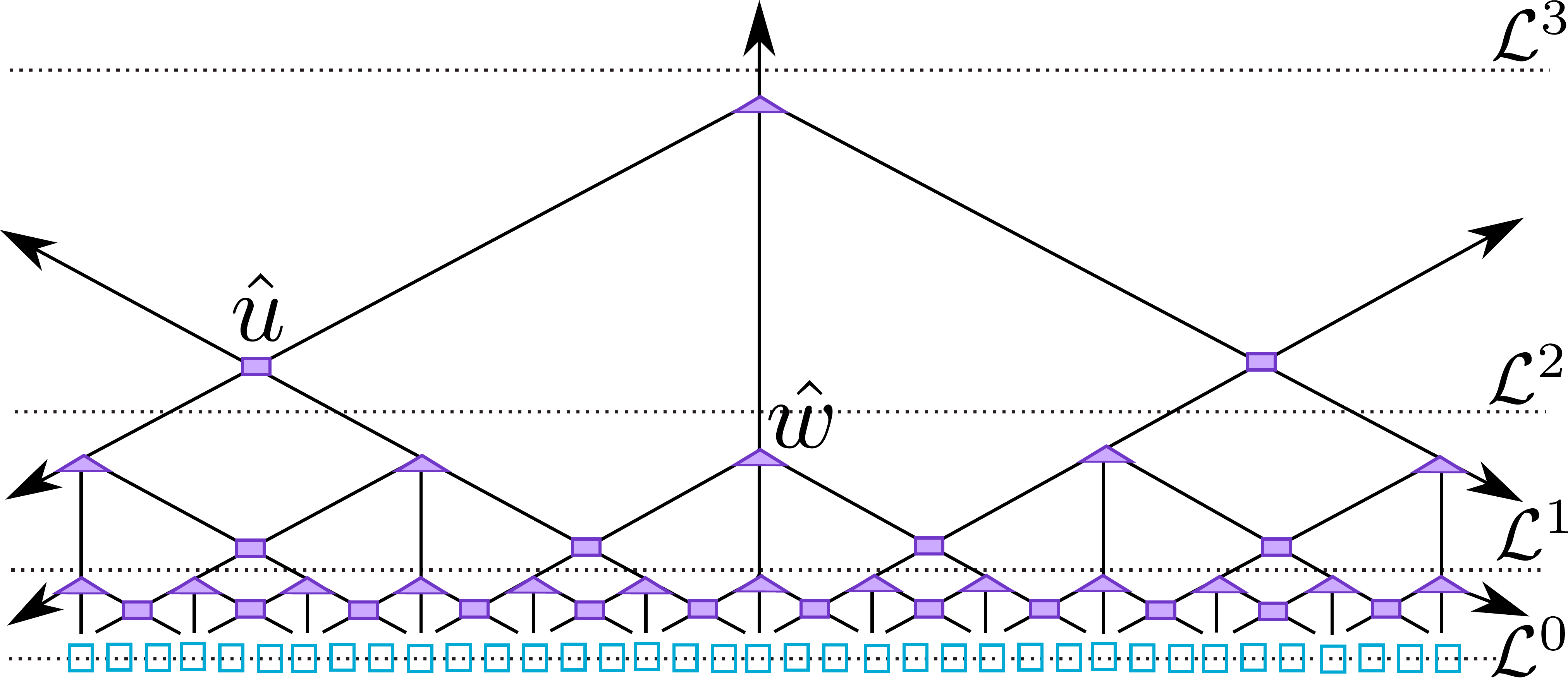}
\caption{\label{fig:mera} (Color online) Graphical representation of the MERA tensor network representation of a quantum many-body state of an infinite lattice $\mathcal{L}$  ($\cong \mathcal{L}_0$). Arrows in the figure indicate that the tensor network extends infinitely in the upward vertical and both left,right horizontal directions. Each open index of the tensor network labels an orthonormal basis on a different site (blue squares) of $\mathcal{L}$. The dotted horizontal lines separate the tensor network into layers of tensors, and coincide with a sequence of increasing coarse-grained lattices: $\mathcal{L}^0 \rightarrow\mathcal{L}^1\rightarrow  \mathcal{L}^2 \cdots$. The vertical direction of the tensor network corresponds to length scale, namely, after discarding the bottom layers the residual tensor network describes the many-body state at a coarser length scale.}
\end{figure}

Consider an infinite 1D lattice $\mathcal{L}$, each site of which is described by a $\chi$-dimensional Hibert space $\mathbb{V}$. Lattice $\mathcal{L}$ is equipped with the action of a local, translational invariant Hamiltonian $\hat{H}$, which may be gapped or critical (gapless). We are interested in the ground state $\mypsi$ of $\hat{H}$. The superscript `bound' appears in anticipation that the ground state will play the role of the boundary state in our holographic correspondence. In this paper, we represent $\mypsi$ by means of an infinite MERA tensor network $\mathcal{T}_{\tiny{\mbox{MERA}}}$, depicted in \fref{fig:mera}. The tensor network $\mathcal{T}_{\tiny{\mbox{MERA}}}$ formally encodes the ground state $\mypsi$ according to \eref{eq:open}, here applied on the infinit lattice $\mathcal{L}$. The MERA representation of the ground state of a given local Hamiltonian can be obtained by means of e.g. the variational energy minimization algorithm \cite{MERAAlgo}.


The MERA representation also describes the RG flow of the ground state. Each layer of MERA tensors, separated by dotted lines in \fref{fig:mera}, implements a real space RG transformation---known as \textit{entanglement renormalization}---that maps a lattice $\mathcal{L}^{k}$ with $L~(\rightarrow \infty)$ sites to a coarse-grained lattice $\mathcal{L}^{k+1}$ with $L/3$ sites.
The MERA tensors are chosen so that the RG transformation preserves the ground subspace at each step. Subsequent RG steps generate a sequence of increasingly coarse-grained lattices: $\mathcal{L}^{0} \rightarrow\mathcal{L}^{1}\rightarrow \mathcal{L}^{2} \cdots$ where $\mathcal{L}^{0} \cong \mathcal{L}$ is the ultraviolet lattice . Therefore, the extra dimension of the tensor network corresponds to length scale. In particular, the residual tensor network obtained after discarding one or more bottom layers of the MERA furnishes a representation of the ground state on a coarse-grained lattice. 

If, in addition to translation invariance, the ground state is also \textit{scale-invariant}---namely, it remains invariant under the RG (entanglement renormalization) transformations---then its MERA representation may be composed of copies of the same two tensors $\hat{u}$ and $\hat{w}$ throughout the tensor network \cite{MERAcrit}. This leads to a very compact description of the infinite ground state, namely, the entire state is completely specified by the two tensors $\hat{u}$ and $\hat{w}$. In the rest of the paper, we will assume that the Hamiltonian $\hat{H}$ (and its ground state $\mypsi$) is also scale-invariant, and that the MERA representation of the ground state is composed of copies of tensors $\hat{u}$ and $\hat{w}$.

The MERA tensors $\hat{u}$ and $\hat{w}$ are constrained to be isometries satisfying:
\begin{equation}\label{eq:isometric}
\sum_{kl} (\hat{u})^{ij}_{kl}(\hat{u}^\dagger)^{kl}_{i'j'} = \delta^i_{i'}\delta^j_{j'},~~\sum_{jkl} (\hat{w})^{i}_{jkl}(\hat{w}^\dagger)^{jkl}_{i'} = \delta^i_{i'},
\end{equation}
where $i,j,k,l,i',j' \in \{1,2,\ldots,\chi\}$.
Consequently, the reduced density matrix of any site on the lattice does not depend on all the MERA tensors, but only on a subset of them; this subset of tensors is called the \textit{causal cone} of the site. The number of tensors in the causal cone that are counted at any given length scale is bounded (less than or equal to 3). Furthemore, the reduced density matrix of multiple sites depends only on tensors belonging to the union of the respective one-site causal cones, which merge at a sufficiently large length scale. Thanks to these properties, expectation values can be efficiently computed from an infinite scale-invariant MERA tensor network \cite{MERA,MERAcrit,MERAAlgo}. 

\subsection{Dual bulk state}\label{sec:bulk}


Let us embed the MERA tensor network $\mathcal{T}_{\tiny{\mbox{MERA}}}$, which represents $\mypsi$, in a 2D manifold with a boundary, such that the open indices and bond indices of the tensor network appear at the boundary and in the bulk of the manifold respectively. 
Next, let us apply tensor network state correspondence to the (embedded) MERA in order to construct a 2D hyperbolic lattice $\mathcal{M}$ and a  tensor network bond state $\myphi$ belonging to this lattice $\mathcal{M}$.

\begin{figure}
  \includegraphics[width=\columnwidth]{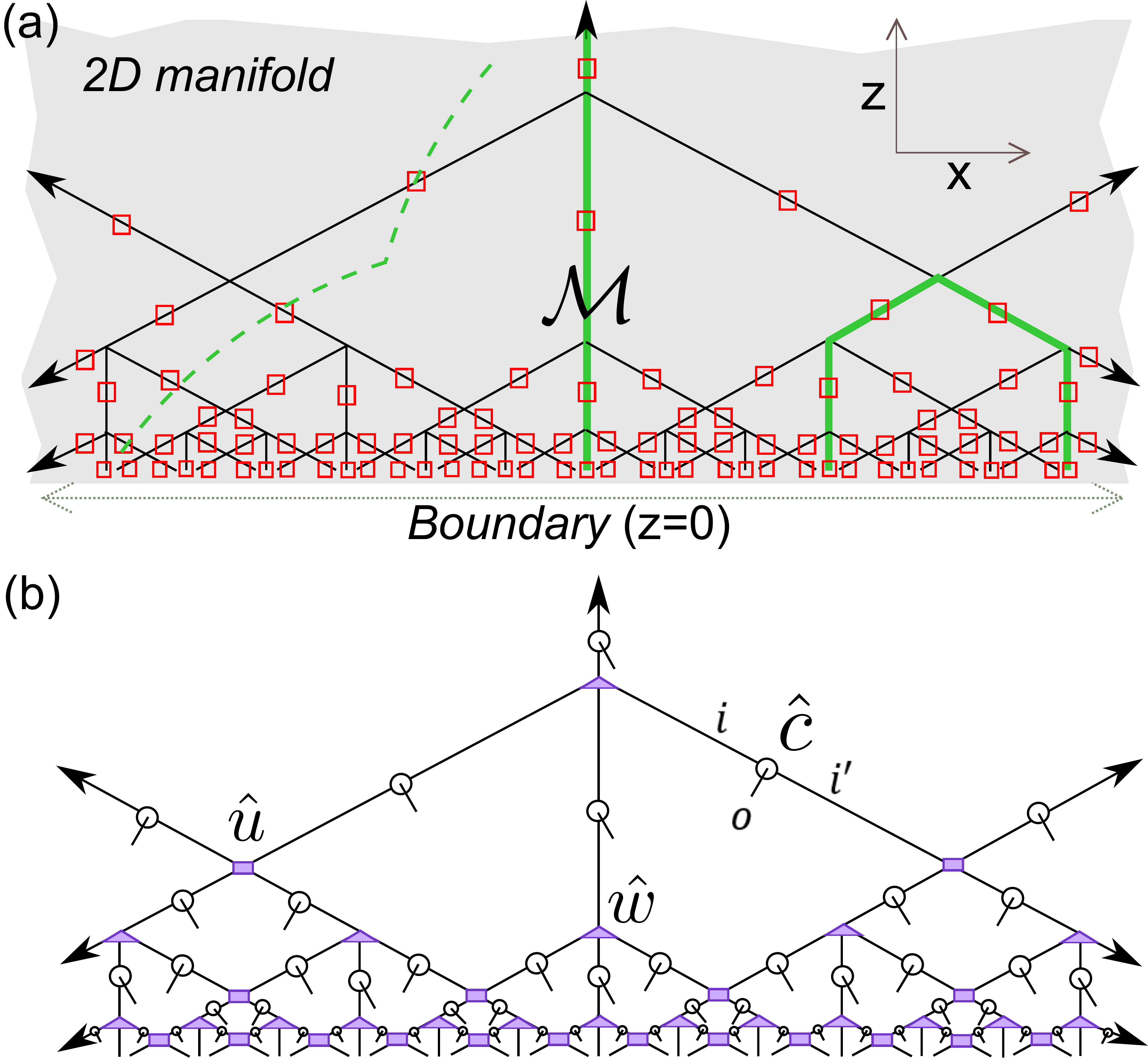}
\caption{\label{fig:lift} (Color online) (a) The dual 2D bulk lattice $\mathcal{M}$, constructed by embedding the MERA in a 2D manifold with a boundary (at $z=0$), and locating a site (red square) of $\mathcal{M}$ on every bond of the MERA. The green solid paths are \textit{graph geodesics}, namely, geodesic paths along the edges of the graph underlying the MERA. Also shown is a path (green dashes) in the ambient manifold that intersects only the edges of the MERA graph. (b) The lifted MERA, our ansatz for the holographic dual state, obtained by inserting a 3-index tensor $(\hat{c})^{i}_{i'o}$ on every bond of the MERA. By using each open index of the tensor network (e.g. index $o$) to label an orthonormal basis on a different site of $\mathcal{M}$, the lifted MERA encodes a quantum state of $\mathcal{M}$.
}
\end{figure}

Construct a 2D quantum lattice $\mathcal{M}$ on the manifold by locating a site---described by the $\chi$-dimensional Hilbert space $\mathbb{V} $---on every bond of the (embedded) tensor network, see \fref{fig:lift}(a). Lattice $\mathcal{M}$ is simply a collation of the degrees of freedom that describe the RG flow of the boundary state, and also inherits the hyperbolic geometry of the tensor network \cite{BulkCount}. 

We then lift the MERA tensor network by inserting a copy tensor [\eref{eq:copytensor}] on each bond index, and use each open index of the lifted MERA to label an orthonormal basis on a different site of $\mathcal{M}$. The lifted MERA encodes a quantum state $\myphi \equiv \bondstate$ of $\mathcal{M}$ according to \eref{eq:bond}. Since we have embedded the MERA in a 2D manifold with a boundary, the two states $\openstate$ and $\bondstate$ are seen to live at the boundary and in the bulk of the hyperbolic lattice $\mathcal{M}$. (Here we identify the sites located at the boundary of $\mathcal{M}$ with the sites of lattice $\mathcal{L}$. For example, the site of $\mathcal{M}$ located at $(x,0)$ is identified with the site of $\mathcal{L}$ located at $x$. Note that the bulk state $\myphi$ also has support on the boundary sites of $\mathcal{M}$.)

The two states $\mypsi$ and $\myphi$constitute our holographic correspondence, mediated by the MERA tensor network $ \mathcal{T}_{\tiny{\mbox{MERA}}}$,
\begin{equation}\label{eq:bbcorresondence}
\mypsi \longleftarrow \mathcal{T}_{\tiny{\mbox{MERA}}} \longrightarrow \myphi.
\end{equation}
For a more general construction, one may also pre-process the MERA by contracting and/or decomposing some of its tensors before lifting it to a 2D quantum state as described here, see Appendix \ref{app:unitaryFreedom}. 

In the remainder of the paper we explore interesting properties of the bulk states constructed from the MERA as described above. We begin by noting two interesting properties of the lifted MERA. First, analogous to the MERA, a lifted MERA is also endowed with a causal cone structure that allows for efficient computation of expectation values in the bulk, as described in Appendix~\ref{app:bulkcone}. Second, the bulk state $\myphi$ described by the lifted MERA contains only a limited entanglement.
Given a subsystem of the bulk lattice $\mathcal{M}$, we define its perimeter and area as the number of sites that lie at the subsystem's boundary and inside the subsystem respectively. For bulk states described by a lifted MERA, the entanglement entropy of a sufficiently large subsystem scales at most as the perimeter of the subsystem, as proved in Appendix \ref{app:arealaw}. Such an entanglement scaling is ubiquitous in condensed matter physics where it is called `(boundary) area law entanglement scaling' and is commonly exhibited by ground states of local, low-dimensional quantum lattice systems \cite{AreaLaw}. In contrast, one expects the subsystem entanglement entropy of generic states of $\mathcal{M}$ to scale as the subsystem's area.

\section{Holographic screens}\label{sec:bulkholo}

\begin{figure}[t]
  \includegraphics[width=\columnwidth]{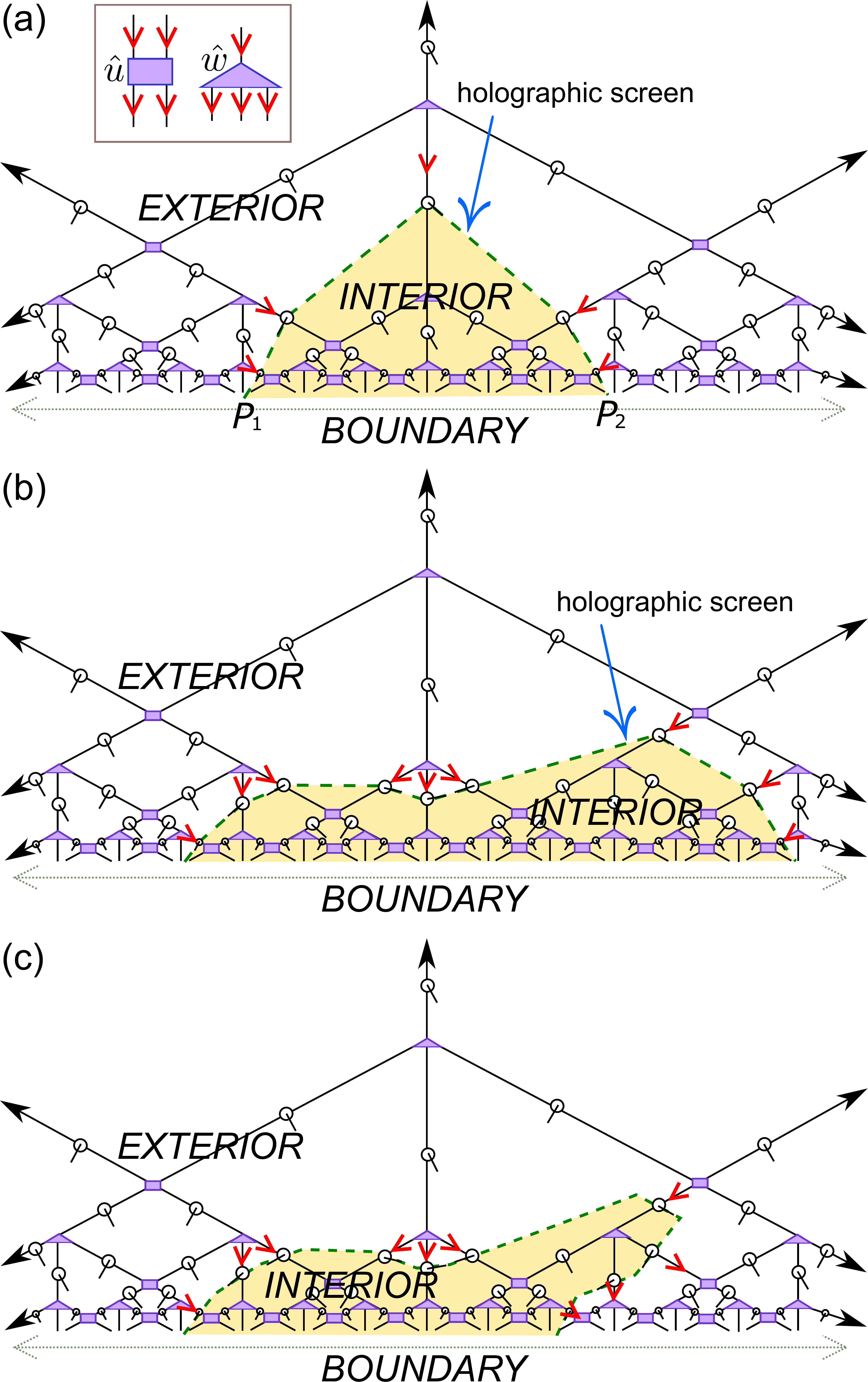}
\caption{\label{fig:holoEnt} (Color online) (a,b) Examples of holographic screens in a lifted MERA.  In the box: a decoration of the indices of the MERA tensors with (red) arrows. Holographic screens are paths on the ambient manifold that extend between two boundary locations (e.g. $P_1, P_2$), intersect only copy tensors, and the red arrows in the immediate exterior of the path are all incoming to the interior. The reduced density matrix of the 2D interior (highlighted in yellow) transforms to the reduced density matrix of the bulk sites located on the (1D) screen under conjugation by an isometry, \eref{eq:holoScreen}. (a) A  geodesic holographic screen (green dashes). (b) A non-geodesic holographic screen (green dashes). (c) Example of a path (green dashes) that \textit{does not} furnish a holographic screen, since some of the red arrows are outgoing from the interior.}
\end{figure}

Let us parameterize the sites of the bulk lattice $\mathcal{M}$ by coordinates $(x,z)$ where, in the boundary description, $x$ labels spatial translations and $z$ corresponds to the length scale (the boundary is located at $z=0$).
Consider two points $P_1$ and $P_2$ at the boundary of the ambient manifold, in which the (lifted) MERA is embedded. $P_1$ is located on the line segment between bulk sites $(x-1,0)$ and $(x,0)$, and $P_2$ is located on the line segment between bulk sites $(x',0)$ and $(x'+1,0)$, for some $x \neq x'$. Consider a path $\mathcal{P}$ between points $P_1$ and $P_2$ that intersects only the copy tensors of the lifted MERA, as illustrated (green dashes) in \fref{fig:holoEnt}. Path $\mathcal{P}$ divides the bulk lattice $\mathcal{M}$ into three parts:
\begin{enumerate}
\item an `interior' composed of bulk sites enclosed between the path and the boundary, and including sites located at $(x,0),(x+1,0), \ldots,(x',0)$,
\item bulk sites associated with the copy tensors that are intersected by $\mathcal{P}$, and
\item an `exterior' composed of all remaining bulk sites.
\end{enumerate}
Let us decorate the indices of tensors $\hat{u}$ and $\hat{w}$ with arrows as depicted (red) in \fref{fig:holoEnt}(a). 
If the arrows on all the bond indices located in the immediate exterior of the path are incoming to the interior then $\mathcal{P}$ can be viewed as a \textit{holographic screen}. It can be shown that $\mathcal{P}$ satisfies (see Appendix  \ref{app:holo})
\begin{equation}\label{eq:holoScreen}
\hat{\rho}^{\tiny \mbox{interior}} = \hat{R}^\dagger (\hat{\rho}^{\tiny \mbox{screen}}) \hat{R},
\end{equation}
where $\hat{\rho}^{\tiny \mbox{interior}}$ is the reduced density matrix of the bulk sites located in the interior, $\hat{\rho}^{\tiny \mbox{screen}}$  is the reduced density matrix of the bulk sites intersected by the screen, and $\hat{R}$ is an isometry, namely, $\hat{R}\hat{R}^\dagger = \hat{I}$. ($\hat{R}$ is obtained by contracting all the tensors in the interior.)

Equation \ref{eq:holoScreen} implies that the expectation value of an observable $\hat{o}^{\tiny \mbox{interior}}$ acting in the interior is equal to the expectation value of the observable $\hat{o}^{\tiny \mbox{screen}} =  \hat{R} (\hat{o}^{\tiny \mbox{interior}}) \hat{R}^\dagger$ acting on the screen, since
\begin{equation} \label{eq:hologram}
\begin{split}
\mbox{Tr}(\hat{\rho}^{\tiny \mbox{interior}}\hat{o}^{\tiny \mbox{interior}}) &= \mbox{Tr}(\hat{R}^\dagger \hat{\rho}^{\tiny \mbox{screen}} \hat{R} \hat{o}^{\tiny \mbox{interior}}) \\
&= \mbox{Tr}( \hat{\rho}^{\tiny \mbox{screen}} \hat{R} \hat{o}^{\tiny \mbox{interior}}\hat{R}^\dagger),\\
&=\mbox{Tr}(\hat{\rho}^{\tiny \mbox{screen}}\hat{o}^{\tiny \mbox{screen}}).
\end{split}
\end{equation}
Here we used \eref{eq:holoScreen} and the cyclic property of trace: $\mbox{Tr}(AB) = \mbox{Tr}(BA)$. 

Thus, the expectation value of any observable supported in the 2D interior region can be calculated from the 1D screen, which encloses the interior. Furthemore, the expectation value of a \textit{local} interior observable equates to the expectation value of a \textit{local} screen observable. Namely, an observable supported on a small number of interior sites maps to an observable that is also supported on a small number of screen sites. This is because $\hat{R}$ is a composition of isometries, each of which act on a small number of sites. (In contrast, the expectation value of a local screen observable $\hat{\omega}^{\tiny \mbox{screen}}$ generally equates to the expectation value of an interior observable  $\hat{\omega}^{\tiny \mbox{interior}} = (\hat{R}^\dagger \hat{R}) \hat{\omega}^{\tiny \mbox{screen}} (\hat{R}^\dagger \hat{R})$ that is smeared over all the interior degrees of freedom.)
From \eref{eq:holoScreen} it also follows that $\hat{\rho}^{\tiny \mbox{screen}}$ and $\hat{\rho}^{\tiny \mbox{interior}}$ have the same eigenvalues, which in turn implies e.g. that the entanglement entropy of all the interior sites \textit{is equal} to the entanglement entropy of all the screen sites, namely,
\begin{equation}\label{eq:holoentropy}
-\mbox{Tr}(\hat{\rho}^{\tiny \mbox{interior}} \mbox{log}_2~\hat{\rho}^{\tiny \mbox{interior}}) = -\mbox{Tr}(\hat{\rho}^{\tiny \mbox{screen}} \mbox{log}_2~\hat{\rho}^{\tiny \mbox{screen}}).
\end{equation}
This entanglement feature is compatible, but extends beyond, the entanglement scaling proved in Appendix \ref{app:arealaw}.

Let us define the length of path $\mathcal{P}$ as the number of copy tensors that it intersects. It can be easily verified that if $\mathcal{P}$ is a geodesic between the points $P_1$ and $P_2$, namely, $\mathcal{P}$ intersects the smallest possible number of copy tensors then it is necessarily a holographic screen [\fref{fig:holoEnt}(a)]. This is because a geodesic path always fulfills the arrow criterion stated previously and therefore satisfies \eref{eq:holoScreen}. On the other hand, \fref{fig:holoEnt}(b) illustrates a non-geodesic holographic screen.

The presence of holographic screens described here is a generic property of the bulk state, and seems to imitate the holographic screens---a feature of (quantum) spacetime---that often appear in quantum gravity.
We also remark that the MERA tensors located in the interior of a holographic screen considered here compose a \textit{local conformal transformation} on the boundary state \cite{localScaleMERA}. Thus, the projection from a bulk region to an enclosing holographic screen may be viewed as the bulk dual of the action of a local conformal transformations on the boundary state.

\section{A bulk/boundary dictionary}\label{sec:dictionary}

\begin{figure}
  \includegraphics[width=\columnwidth]{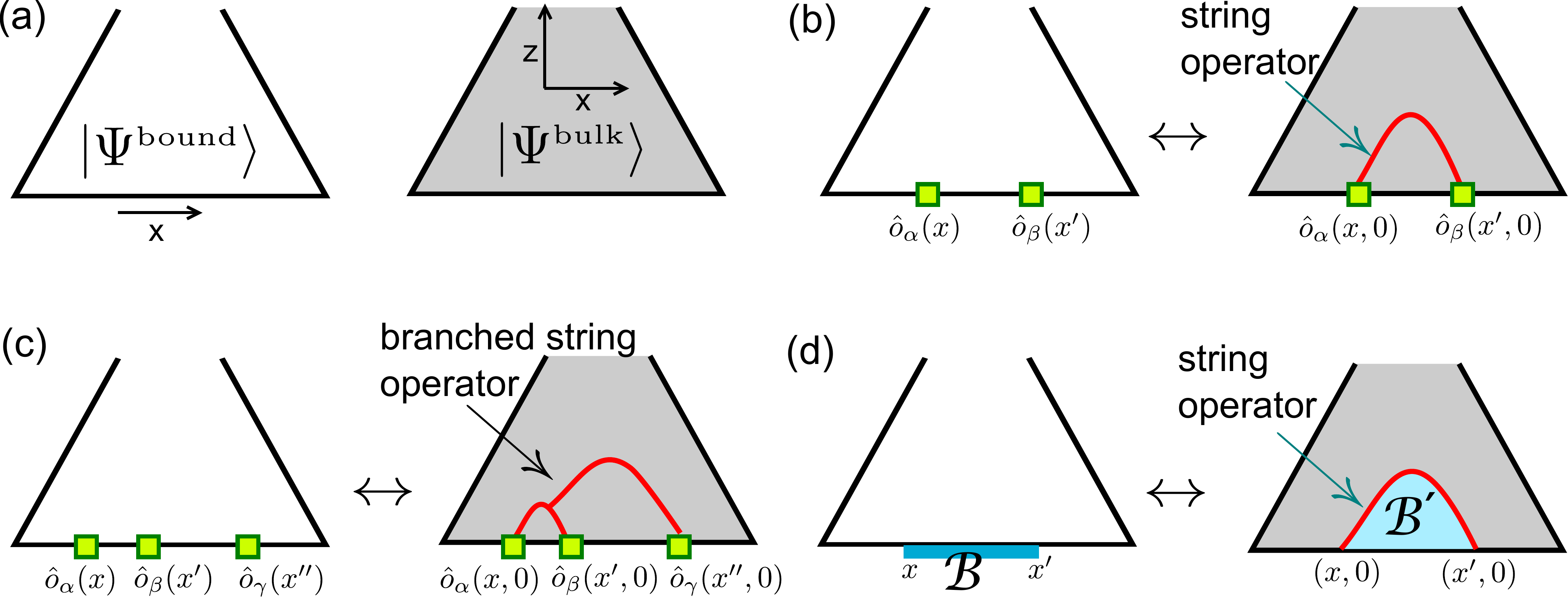}
\caption{\label{fig:dictionary} (Color online) A dictionary that translates between properties of a critical ground state $\mypsi$, represented by a MERA, and the dual bulk state $\myphi$. Here $x,x',x''$ locate special sites on the lattice $\mathcal{L}$, see main text. (a) A schematic depiction of the MERA (represents $\mypsi$) and a lifted MERA (represents $\myphi$). (b) The 2-point boundary correlator $\xbound{\hat{o}_\alpha({x}) \hat{o}_\beta({x'})}$ is proportional to the bulk expectation value of a string operator, \eref{eq:corr}.   (c) The 3-point correlator $\xbound{\hat{o}_\alpha({x}) \hat{o}_\beta({x'}) \hat{o}_\gamma({x''})}$ is proportional to the bulk expectation value of a branched string operator, \eref{eq:OPE}. (d) The Reyni entropy of a boundary region $\mathcal{B} \subset \mathcal{L}$ is equal to the Reyni entropy (plus a constant) of a bulk region $\mathcal{B}'\subset \mathcal{M}$ as calculated from the projected bulk state $\ket{\Omega^{\tiny \mbox{bulk}}}$, \eref{eq:reyni}. $\mathcal{B}'$ is comprised of sites enclosed between the geodesic that extends between $x$ and $x'$ and the boundary, including the boundary sites of $\mathcal{M}$ located at $(x+1,0),(x+2,0),\ldots,(x'-1,0)$.}
\end{figure}

In this section, we describe a simple prescription to obtain correlators and block Reyni entanglement entropy of a \textit{critical} ground state $\mypsi$ represented by a MERA tensor network from the dual bulk state $\myphi$ obtained from the tensor network. This prescription identifies correlators and block Reyni entanglement entropy of the critical ground state to corresponding quantities in the bulk states, and illustrates that the properties of the bulk and boundary state are indeed systematically related together. The prescription is schematically depicted in \fref{fig:dictionary}. For simpler presentation, here we only list the formulae for calculating these boundary properties from the bulk, while their derivation is presented in Appendix \ref{app:critical}. 

For the purpose of this section, we denote by $x,x',x''$ the locations of special sites of the boundary lattice $\mathcal{L}$, namely, $x$ locates a site associated with an open index of the MERA at the base of an arbitrarily long vertical graph geodesic [see \fref{fig:lift}(a)], $|x-x''| = 3^q$, and $|x'-x'|=3^{q'}$ where $q,q'$ are positive integers. Here by `graph geodesic' we mean the shortest connected path between two locations along the MERA graph itself, in contrast with geodesics on the ambient manifold outside the lattice that were considered in the previous section, Sec. \ref{sec:bulkholo}. Let $\mathcal{G}_{x,x'}$ denote the set of bulk sites that are located along the graph geodesic extending between the bulk sites at $(x,0)$ and $(x',0)$.

Consider the \textit{one-site scaling superoperator} $\hat{S}$ obtained from the MERA tensor $\hat{w}$ as 
\begin{equation}
(\hat{S}^i_{i'})^{k'}_ {k} = \sum_{jl} \hat{w}^{i}_{jkl} (\hat{w}^\dagger)^{jk'l}_{i'},\nonumber
\end{equation}
and let $\hat{o}$ and $\lambda$ denote an eigenoperator and the corresponding eigenvalue of $\hat{S}$, namely, $\hat{S}(\hat{o}) = \lambda \hat{o}$.
Operator $\hat{o}$ is identified with a scaling operator of the underlying CFT with scaling dimension $\Delta = -\mbox{log}_3 \lambda$ \cite{MERAcrit}. 
The \textit{2-point boundary correlator} of scaling operators $\hat{o}_\alpha$ and $\hat{o}_\beta$, applied at site locations $x$ and $x'$, can be obtained from the bulk state as
\begin{equation}\label{eq:corr}
\begin{split}
\xbound{\hat{o}_\alpha({x}) \hat{o}_\beta({x'})}  \propto \xbulk{\hat{K}_{x,x'}\hat{o}_\alpha({x,0}) \hat{o}_\beta({x',0})},
\end{split}
\end{equation}
where $\hat{K}_{x,x'} = \bigotimes_{i}\hat{P}_{i}$ is a \textit{string operator} that projects each bulk site $i \in \mathcal{G}_{x,x'}$ to the state $\ket{+}$ [\eref{eq:recoverstate}]. Equivalently, the boundary correlator may be obtained by calculating the same correlator from the bulk state but after projecting all the bulk sites in $\mathcal{G}_{x,x'}$ to the state $\ket{+}$.

Analogously, the \textit{3-point boundary correlator} of scaling operators $\hat{o}_\alpha, \hat{o}_\beta$ and $\hat{o}_\gamma$ can be obtained from the bulk state as
\begin{equation}\label{eq:OPE}
\begin{split}
&\xbound{\hat{o}_\alpha(x) \hat{o}_\beta({x'}) \hat{o}_\gamma({x''})} \propto \\
&\xbulk{\hat{T}_{x,x',x''} \hat{o}_\alpha(x,0) \hat{o}_\beta({x',0}) \hat{o}_\gamma({x'',0})},
\end{split}
\end{equation}
where $\hat{T}_{x,x',x''} = \bigotimes_{i}\hat{P}_{i}$ is a branched string operator that projects all bulk sites $i \in \{\mathcal{G}_{x,x'} \cup \mathcal{G}_{x',x''}\}$ to the  state $\ket{+}$.

More generally, the $n$-point correlator $\xbound{\hat{o}_\alpha(x_1),\hat{o}_\beta({x_2}),\cdots,\hat{o}_\nu(x_n)}$ is proportional to the bulk expectation value of the $n$ scaling operators tensor product with \textit{multi-branched} string projection operator, analogous to $\hat{T}_{x,x',x''}$ but with $n-2$ branch points. (Once again, $x_i$'s locate special sites of $\mathcal{L}$, as described above.)
Thus, $n$-\textit{point} correlators of an 1D critical ground state translate to the expectation value of \textit{extended} operators in a dual 2D bulk state.
This identification, though extremely simple, appears to caricature the prescription of pertubatively calculating boundary correlators in AdS/CFT by evaluating Witten diagrams, as stated in the introduction. (The latter is a type of Feyman diagram that involves integrating certain bulk degrees of freedom and propagating the boundary operators into the bulk. Note in comparison here that applying the projector $\hat{P}$, \eref{eq:recoverstate}, on a bulk site in effect sums the basis vectors at the site.)

Note that the extended operators that appear in \eref{eq:corr} and \eref{eq:OPE} act on a \textit{finite} number of bulk sites. On the other hand, any of the boundary correlators considered above can be obtained \textit{exactly} by computing the same correlator in the bulk state but after projecting an \textit{infinite} number of bulk sites---those located in the joint causal cone of the sites on which the scaling operators act---to the state $\ket{+}$. (This can be seen by matching the tensor network contraction equating to the boundary correlator with the tensor network contraction equating to the same correlator in the bulk after projecting all the bulk sites in the joint causal cone.)  The point here is that in order to obtain the critical exponents, which determine the scaling of the boundary correlators and are part of the underlying CFT data, from the bulk it suffices to consider the bulk expectation value of extended operators that act only on a finite number of bulk sites.

The Reyni entanglement entropy of a block $B$ of sites in a quantum many-body state $\ket{\Psi}$ is:
\begin{equation}
R_{\alpha}(B) = -\mbox{log}_2 \mbox{ Tr} \rhoy_B^{\alpha},~~~\alpha = 2,3,\ldots,\nonumber
\end{equation}
where $\rhoy_B$ is the reduced density matrix of the block $B$, obtained by tracing out all sites belonging to the complement of $B$ in state $\ket{\Psi}$.
The boundary Reyni entanglement entropy $R_{\alpha}^{\tiny \mbox{bound}}(\mathcal{B})$ of a block $\mathcal{B} \subset \mathcal{L}$ of sites located at $x,x+1,\ldots,x'$ can be also be obtained from the bulk in a simple way in the limit of large block size $|x-x'|$. We have
\begin{equation}\label{eq:reyni}
\begin{split}
R_{\alpha}^{\tiny \mbox{bound}}(\mathcal{B}) \approx
R_{\alpha}^{\tiny \mbox{bulk}}(\mathcal{B}') + \mbox{const},~~~|x-x'| \gg 1,
\end{split}
\end{equation}
where (i) $\mathcal{B}' \subset \mathcal{M}$ [see \fref{fig:dictionary}(d)] is the set of bulk sites enclosed between the graph geodesic extending between locations $(x,0)$ and $(x',0)$ and the boundary, and including the bulk sites located at $(x+1,0),(x+2,0),\ldots,(x'-1,0)$, (ii) $R_{\alpha}^{\tiny \mbox{bulk}}(\mathcal{B}')$ is the Reyni entanglement entropy of $\mathcal{B}'$ calculated from the \textit{projected bulk state} $\ket{\Omega^{\tiny \mbox{bulk}}} = \hat{K}_{x,x'}\myphi$, and (iii) the additive constant does not depend on the size $|x-x'|$ of the block. 

Note that \eref{eq:reyni} differs from the equality \eref{eq:holoentropy} between the entanglement entropy of sites located on and inside a holographic screen respectively. While \eref{eq:reyni} relates a \textit{boundary} Reyni entanglement entropy to a \textit{bulk} Reyni entanglement entropy and holds approximately in the limit of large $|x-x'|$, \eref{eq:holoentropy} is an exact property of the bulk state.

\section{Bulk entanglement vs boundary central charge}\label{sec:bulkentanglement}

As stated in the introduction, certain bulk features in the AdS/CFT correspondence depend on the central charge of the CFT. For example, correlations due to quantum fluctuations in the bulk are generally suppressed for large central charge \cite{quantumFlucBulk}. Another example is the Ryu-Takayanagi formula, which holds when the bulk is described by classical gravity \cite{HoloEntropy}. The Ryu-Takayanagi formula equates (in appropriate units) the von Neumann entanglement entropy $S^{\tiny \mbox{bound}}_\ell$ of an interval of length $\ell$ in the CFT vacuum to the length $L_{\tiny \mbox{geo}}$ of the geodesic that extends between the end points of the interval through the dual bulk spacetime, $S^{\tiny \mbox{bound}}_\ell \approx L_{\tiny \mbox{geo}}$. We also have $S^{\tiny \mbox{bound}}_\ell = \frac{c}{3}~\mbox{log} \ell$, where $c$ is the central charge \cite{ReyniCFT}. Thus, $S^{\tiny \mbox{bound}}_\ell$, and therefore $L_{\tiny \mbox{geo}}$ determined in the dual bulk geometry, increases with the central charge $c$.

Motivated by these features, it would be interesting to probe for any potential dependence of the bulk entanglement/correlations on the boundary central charge. However, in order to compare bulk properties corresponding to different boundary states, one has to first address the ambiguity in defining the bulk state dual to a given ground state that arises due to the `gauge freedom' inherent in tensor network representations.

Given a MERA representation of a quantum many-body state, one can obtain another equivalent MERA representation of the state by inserting a resolution of identity $\hat{M}_k\hat{M}_k^{-1}$ on bond $k$, and multiplying the matrices $\hat{M}_k$ and $\hat{M}_k^{-1}$ respectively with the two tensors that are connected by the bond. This \textit{bond freedom} is an intrinsic property of tensor network representations of quantum many-body states. The two MERAs are an equivalent representation of the state since the expectation value of any observable is equal in both representations. (Obtaining an expectation value from a tensor network state involves contracting all the bond indices of the tensor network, and $\hat{M}_k$ is multiplied with $\hat{M}_k^{-1}$ in the process.)

Clearly, inserting the copy tensor defined in \eref{eq:copytensor} selects out a particular MERA representation of the ground state---the one whose tensors are expressed in the basis in which the copy tensor has the components of \eref{eq:copytensor}. Consider bulk states $\ket{\Psi}$ and $\ket{\Psi'}$ that are obtained by lifting MERA tensor networks $\mathcal{T}$ and  $\mathcal{T}'$ respectively. Here MERA $\mathcal{T}'$ is obtained by transforming the tensors of $\mathcal{T}$ by means of \textit{non-diagonal} unitary bond transformations ($\hat{M}_k$'s) as described above. In this case, even though the two MERAs $\mathcal{T}$ and $\mathcal{T}'$ describe the same quantum many-body state, the two corresponding bulk states $\ket{\Psi}$ and $\ket{\Psi'}$ are generally different e.g. they have different entanglement. This is because the copy tensor $\hat{c}$ `commutes' with only \textit{diagonal} matrices, namely, a contraction of $\hat{c}$ with a diagonal matrix on any index is equal to a contraction of $\hat{c}$ with the same diagonal matrix on a different index. This implies that $\ket{\Psi}$ and $\ket{\Psi'}$ are not related to each other by one-site unitary rotations on the bulk lattice, and therefore they have different entanglement.
Thus, our bulk construction generally lifts a given ground state to a set of different bulk states.

In such a scenario, how can one compare bulk states corresponding to different boundary states? For example, one could be interested in probing if the bulk entanglement depends on the boundary central charge, see Sec.~\ref{sec:bulkentanglement}. A possible approach to compare bulk states corresponding to different boundary states is to average the bulk properties over all the different bulk states that are dual to the ground state. Another possibility is to compare the \textit{statistics} of the bulk properties by randomly sampling from the different bulk states. see Ref.~\onlinecite{TNCSym}. 
On the other hand, one may take the view that a given ground state must lift to only one dual bulk state to begin with. A possible way to achieve this, within the framework introduced thus far, is to constrain the intrinsic bond freedom in MERA representations by demanding that the tensors fulfill additional constraints.

For example, in Ref.~\onlinecite{TNCSym} we describe how the fomalism presented in this paper can be extended to implement the holographic translation of a boundary on-site global symmetry to a bulk local gauge symmetry. This is achieved by partially constraining the bond freedom, which illustrates that constraining the bond freedom may indeed be useful (or even necessary) to implement certain features of the AdS/CFT in our holographic correspondence.
In Appendix \ref{app:unitaryFreedom} we introduce two modified MERA representations that have a significantly constrained bond freedom, as compared to the standard MERA representation which is reviewed in Sec.~\ref{sec:boundary}. One of these constrained MERA representations was used to obtain the numerical results presented in Sec.~\ref{sec:bulkentanglement}. However, these constrained MERA representations are not directly motivated from AdS/CFT. Nonetheless, they illustrate a possible generalization of our bulk construction to control the number of different bulk states dual to a given ground state.

It is also possible that the different bulk states, dual to a ground state, obtained here are related to one another by an unidentified (emergent) bulk symmetry, and are thus equivalent holographic duals of the ground state. Or that the relevant dual bulk state is a certain superposition of the different bulk states that is determined by some holography inspired bulk conditions. We leave further exploration of these issues for future work.

\begin{figure}
  \includegraphics[width=7.5cm]{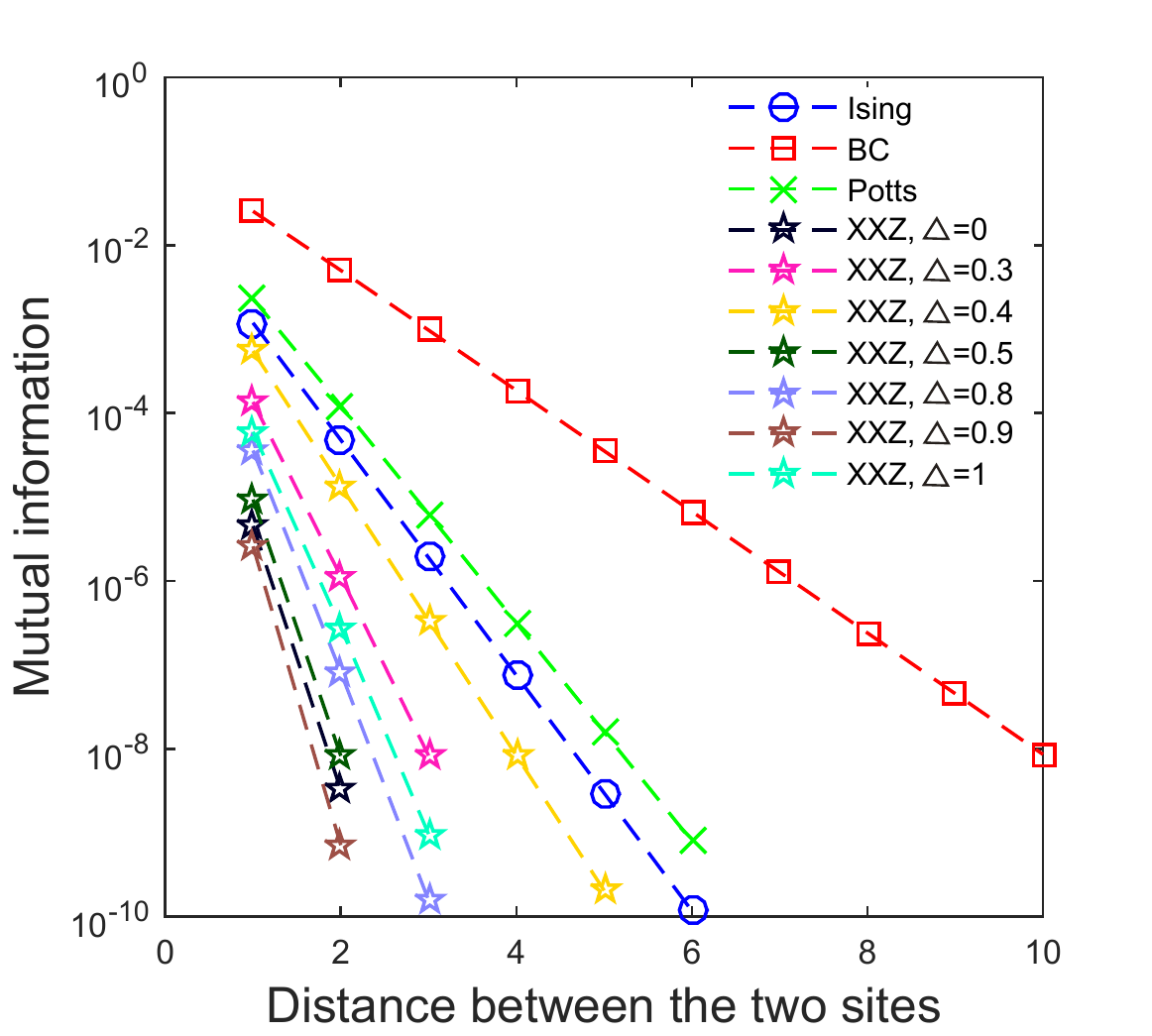}
\caption{\label{fig:results1} (Color online) The mutual information $I^{\tiny \mbox{bulk}}_{z,z'}$ [\eref{eq:mutualResults}] in bulk states, dual to the ground state of each of the critical spin chains listed in \eref{eq:models}, between two bulk sites located at $(x,z)$ and $(x,z')$ respectively (separated by distance $|z-z'|$) as illustrated in \fref{fig:canonical}(c). }
\end{figure}

With this discussion in mind, let us proceed to compare bulk entanglement corresponding boundary states. To this end, we considered the ground states of the following 1D critical spin models with different central charges:
\begin{equation}\label{eq:models}
\begin{split}
\hat{H}^{\mbox{\small \textsc{ising}}} &= \sum_{i} \hat{\sigma}^i_x\hat{\sigma}^{i+1}_x + \hat{\sigma}^i_z,\\
\hat{H}^{\mbox{\small \textsc{bc}}} &= \sum_{i} -\hat{S}_x^i\hat{S}_x^{i+1} + \alpha (\hat{S}_x^i)^2 + \beta \hat{S}_z^i,\\
\hat{H}^{\mbox{\small \textsc{potts}}} &= -\sum_{i} \hat{P}^i(\hat{P}^{T})^{i+1} + (\hat{P}^{T})^{i}\hat{P}^{i+1} + \hat{M}^i,\\
\hat{H}^{\mbox{\small \textsc{xxz}}} &= \sum_{i} \hat{\sigma}^i_x\hat{\sigma}^{i+1}_x + \hat{\sigma}^i_y\hat{\sigma}^{i+1}_y + \Delta \hat{\sigma}^i_z\hat{\sigma}^{i+1}_z,
\end{split}
\end{equation}
where $i$ labels sites of an 1D lattice on which the Hamiltonian acts, $\hat{\sigma}_x,\hat{\sigma}_y,\hat{\sigma}_z$ are Pauli matrices, the operator $\hat{S}_{\alpha}$ is the $\alpha$ component of the spin$-1$ representation of $\mathfrak{su}(2)$, and $\hat{P},\hat{M}$ are $3 \times 3$ Potts matrices:
\[
\hat{P}=\left(\begin{array}{ccc}0 & 1 & 0 \\0 & 0 & 1  \\1 & 0 & 0\end{array}\right);\quad \hat{M}=\left(\begin{array}{ccc}2 & 0 & 0 \\0 & -1& 0 \\0 & 0 & -1\end{array}\right)
\]
$\hat{H}^{\mbox{\small \textsc{ising}}}$ has central charge $\frac{1}{2}$, $\hat{H}^{\mbox{\small \textsc{bc}}}$ is the spin-1 Blume Capel model which is critical for $\alpha=0.910207, \beta=0.415685$ with central charge $\frac{7}{10}$, $\hat{H}^{\mbox{\small \textsc{potts}}}$ has central charge $\frac{4}{5}$, and $\hat{H}^{\mbox{\small \textsc{xxz}}}$ is critical for $-1\leq\Delta<1$ with central charge $1$. (The scaling dimensions of the CFT underlying $\hat{H}^{\mbox{\small \textsc{xxz}}}$ vary continuously with $\Delta$.) For the results presented in this section, we considered values of $\Delta \in \{0,0.3,0.4,0.5,0.8,0.9,1\}$.

\begin{figure}
  \includegraphics[width=7.5cm]{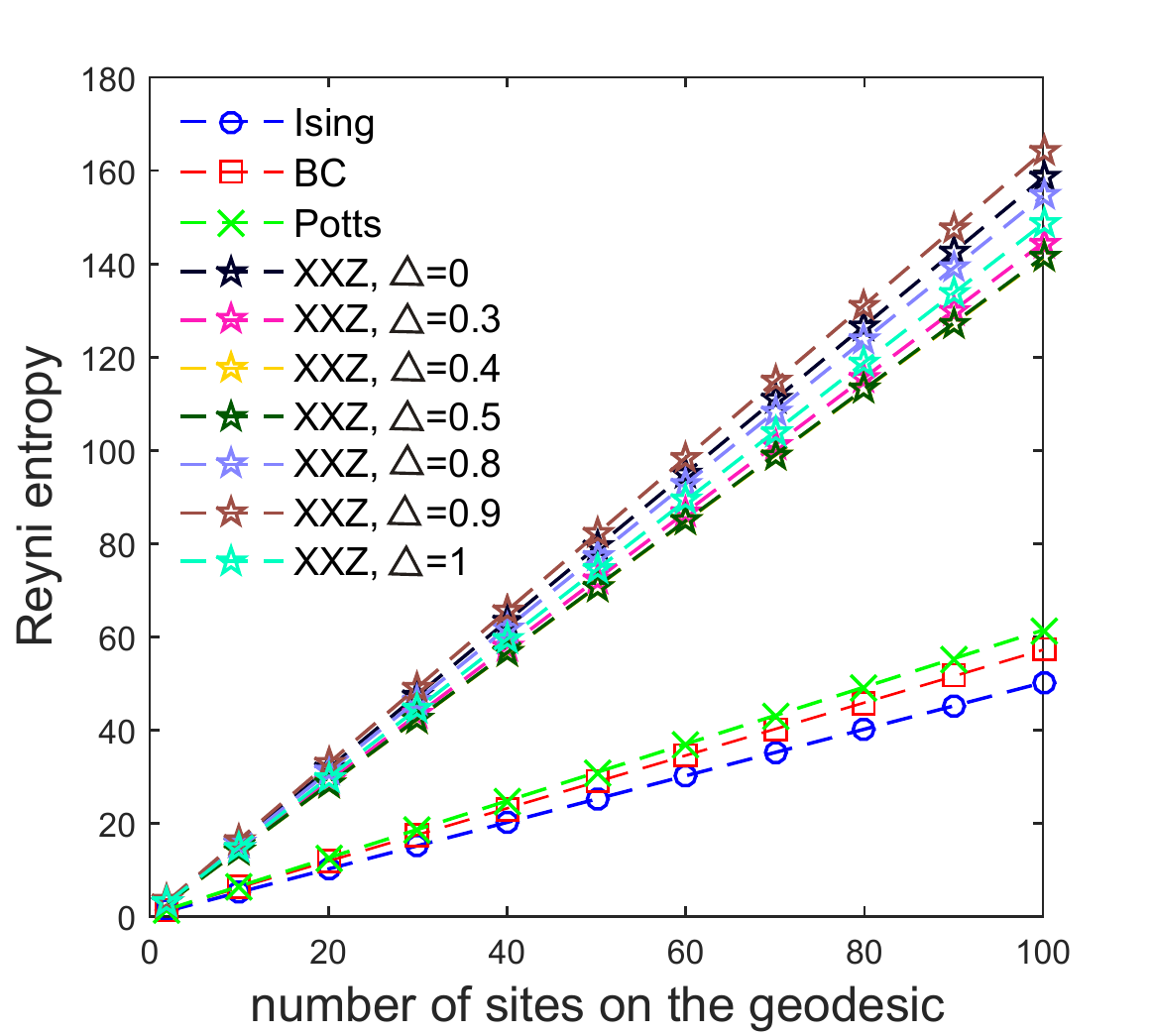}
\caption{\label{fig:results2} (Color online) Second Reyni entanglement entropy in bulk states, dual to the ground state of each of the critical spin chains listed in \eref{eq:models}, of bulk sites located on a geodesic holographic screen [such as the one illustrated in \fref{fig:canonical}(c)].}
\end{figure}

The MERA representation of the ground states of these models was obtained using the variational energy minimization algorithm for the scale-invariant MERA described in Ref.~\onlinecite{MERAcrit} (Pfeifer et al.) keeping bond dimension $\chi = 12$. In these simulations, the error in the ground state energy density for the Ising model was $O(10^{-8})$ while the relative error in the central charge was $0.4\%$. For the remaining models, the error in the ground state energy density was $O(10^{-5})$ and the relative error in the central charge was at most $1.2\%$. The relative error in the first few scaling dimensions for all the models was at most $4\%$.

The Hamiltonians listed in \eref{eq:models} are not scale-invariant, but flow to a scale-invariant fixed point after possibly several RG (entanglement renormalization) steps. We considered the renormalized scale-invariant ground state of each model, described by retaining only the scale-invariant part of the MERA tensor network \cite{MERAcrit}. We first translated the scale-invariant part of the MERA to a constrained MERA form based on higher order singular value decomposition (in order to faciliate the comparison of bulk properties for different ground states). We then lifted the modifed MERA by inserting copy tensors on the bonds to obtain a dual bulk state, as described in Appendix \ref{app:unitaryFreedom}. See discussion in Sec.~\ref{sec:bulk}.

Let us parameterize the bulk lattice $\mathcal{M}$ by coordinates $(x,z)$ where $z$ corresponds to the length scale and $x$ labels spatial translations (in the boundary description). For the purposes of this section, let $(x,z),(x,z')$ and $(x',z)$ locate bulk sites along an arbitrarily long vertical graph geodesic [illustrated in \fref{fig:lift}(a)] such that $z\neq z'$ and $|x-x'| = 3^p$ where $p$ is a positive integer. 

For the ground state of each critical Hamiltonian listed in \eref{eq:models}, we estimated the correlation length  in the dual bulk state along the $z$ direction from the scaling of the mutual information $I^{\tiny \mbox{bulk}}_{z,z'}$ given by
\begin{equation}\label{eq:mutualResults}
I^{\tiny \mbox{bulk}}_{z,z'} = S^{\tiny \mbox{bulk}}_z + S^{\tiny \mbox{bulk}}_{z'} - S^{\tiny \mbox{bulk}}_{z,z'}.
\end{equation}
Here $S^{\tiny \mbox{bulk}}_z = - \mbox{Tr} (\hat{\rho}^{\tiny \mbox{bulk}}_{z}\mbox{log}_2 \hat{\rho}^{\tiny \mbox{bulk}}_{z})$ is the von Neumann entanglement entropy of the bulk site located at $(x,z)$, $S^{\tiny \mbox{bulk}}_{z'}$ is the von Neumann entanglement entropy of the bulk site located at $(x,z')$,  and $S^{\tiny \mbox{bulk}}_{z,z'}$ is the joint von Neumann entanglement entropy of the two sites, see \fref{fig:canonical}(c).

The results are plotted in \fref{fig:results1}. We find that the mutual information decays exponentially, which implies that the bulk state has a finite correlation length along this direction. (The mutual information gives an upper bound for all 2-point correlators.) The plot also suggests a trend that the correlation length is generally smaller for larger boundary central charge for these models (with the exception of the Ising model). On the other hand, as mentioned previously, correlations due to quantum fluctuations in the bulk are also generally suppressed for large central charge in the AdS/CFT correspondence.

We also computed the second Reyni entanglement entropy $R^{\tiny \mbox{bulk}}_{x,x'} = -\mbox{log}_2 \mbox{Tr} (\hat{\rho}^{\tiny \mbox{bulk}}_{x,x'})^2 $. Here $\hat{\rho}^{\tiny \mbox{bulk}}_{x,x'}$ is the reduced density matrix of the bulk sites that are located on a geodesic holographic screen that is anchored next to the bulk sites located at $(x,0)$ and $(x',0)$ [see Sec.~\ref{sec:bulkholo} and also \fref{fig:canonical}(c)] . The results are plotted in \fref{fig:results2}.
We find that the $R^{\tiny \mbox{bulk}}_{x,x'}$ increases linearly with the number of bulk sites located on the screen (proportional to $p$), which is consistent with the area law scaling of bulk entanglement derived in Appendix \ref{app:arealaw}. Interestingly, for any two models with different central charges, the slope of the scaling of $R^{\tiny \mbox{bulk}}_{x,x'}$ is generally larger for the model with larger central charge.

Furthermore, the value of Reyni entanglement entropy $R^{\tiny \mbox{bulk}}_{x,x'}$ for a given block size is larger for a spin chain with a larger central charge. On the other hand, as described at the beginning of this section, in AdS/CFT the length $L_{\tiny \mbox{geo}}$ of the bulk geodesic which extends between the ends of the block also increases with boundary central charge, in accordance with the Ryu-Takayanagi formula. In view of this, it may be interesting to explore whether a bulk metric can be deduced from the bulk entanglement entropy. For example, is the entanglement entropy of the bulk sites intersected by a geodesic, plotted in \fref{fig:results2}, a legitimate measure of the geodesic's length, $L_{\tiny \mbox{geo}} \approx R^{\tiny \mbox{bulk}}_{x,x'}$? We leave this as an open question for future work.
(Computing the von Neumann entanglement entropy, which appears in the Ryu-Takayanagi formula, incurs a higher computational cost. But for short geodesics we verified that the slopes of the scaling of von Neumann entanglement entropy, analogous to $R^{\tiny \mbox{bulk}}_{x,x'}$, also increase monotonically with increase in central charge for these models.)

\section{Generalizations}\label{sec:outlook}

The bulk construction described in this paper can be generalized in several ways. Motivated by further guidelines from holography, one may insert tensors different from the copy tensor on the bonds, or pre-process the tensor network in some way before lifting it (e.g. as discussed in Appendix \ref{app:unitaryFreedom}).
It may also be interesting to consider inserting copy tensors on a fewer number of bonds e.g. only those that are output from the $\hat{w}$ tensors since, strictly speaking, only these are associated with renormalized sites in the MERA, see \fref{fig:mera}. This corresponds to a fewer number of bulk sites. On the other hand, allocating bulk sites to all the bonds, as we have done in this paper, allows the introduction of suitable gauge transformations in the bulk, which are dual to a global on-site symmetry at the boundary \cite{TNCSym}. In the presence of on-site symmetries at the boundary, it is also more natural to associate two bulk sites with every bond of the MERA, which corresponds to lifting the MERA by inserting a 4-index bond tensor with each index taking $\chi$ values. (Or equivalently, a 3-index bond tensor, as in this paper, but whose open index takes $\chi^2$ values.) 

One could also consider attaching additional bulk degrees of freedom with the tensors in the MERA, by internally decomposing each MERA tensor in terms of trivalent tensors, and lifting the internal indices thus exposed. This is more natural in the extended formalism with symmetries presented in Ref.~\cite{TNCSym}, where these additional degrees of freedom can be interpreted as `gauge matter'.

In conclusion, we have tried to make a case for obtaining an emergent 2D description from the MERA representation of an 1D ground state by associating dual degrees of freedom with the bonds of the tensor network. The formalism introduced in this paper illustrates a possible way in which the MERA could implement holography, and more generally how a tensor network with open indices intrinsically encodes a correspondence between two quantum many-body states.

\textbf{Note added.}---After this paper appeared on arXiv.org, a recent paper \cite{GaugeHoloCodes} was brought to my attention that presents a bulk construction that appears some what similar to the one presented here. However, Ref. \onlinecite{GaugeHoloCodes} treats the tensor network as an isometry from a bulk Hilbert space to the boundary Hilbert space. In contrast, here we construct two separate tensor network states---the boundary state (represented by a MERA) and the corresponding bulk state (represented by a lifted MERA)---which are not necessarily related together by an isometry, but instead share a set of tensors.
Furthermore, the focus and subsequent applications of the bulk construction are substantially different in the two papers.

\textbf{Acknowledgements.}---Most of this research was completed while SS was employed at the Australian Research Council's Center of Exellence for Engineered Quantum Systems in Macquarie University. SS thanks Gavin Brennen for many important discussions, and also Guifre Vidal, Juan Maldacena, Nathan McMahan, and Giandemenico Palumbo for stimulating discussions. SS also acknowledges the hospitality of the Perimeter Institute for Theoretical Physics where a part of this work was presented.


\appendix

\section{Causal cone structure of the lifted MERA}\label{app:bulkcone}

\begin{figure}
  \includegraphics[width=\columnwidth]{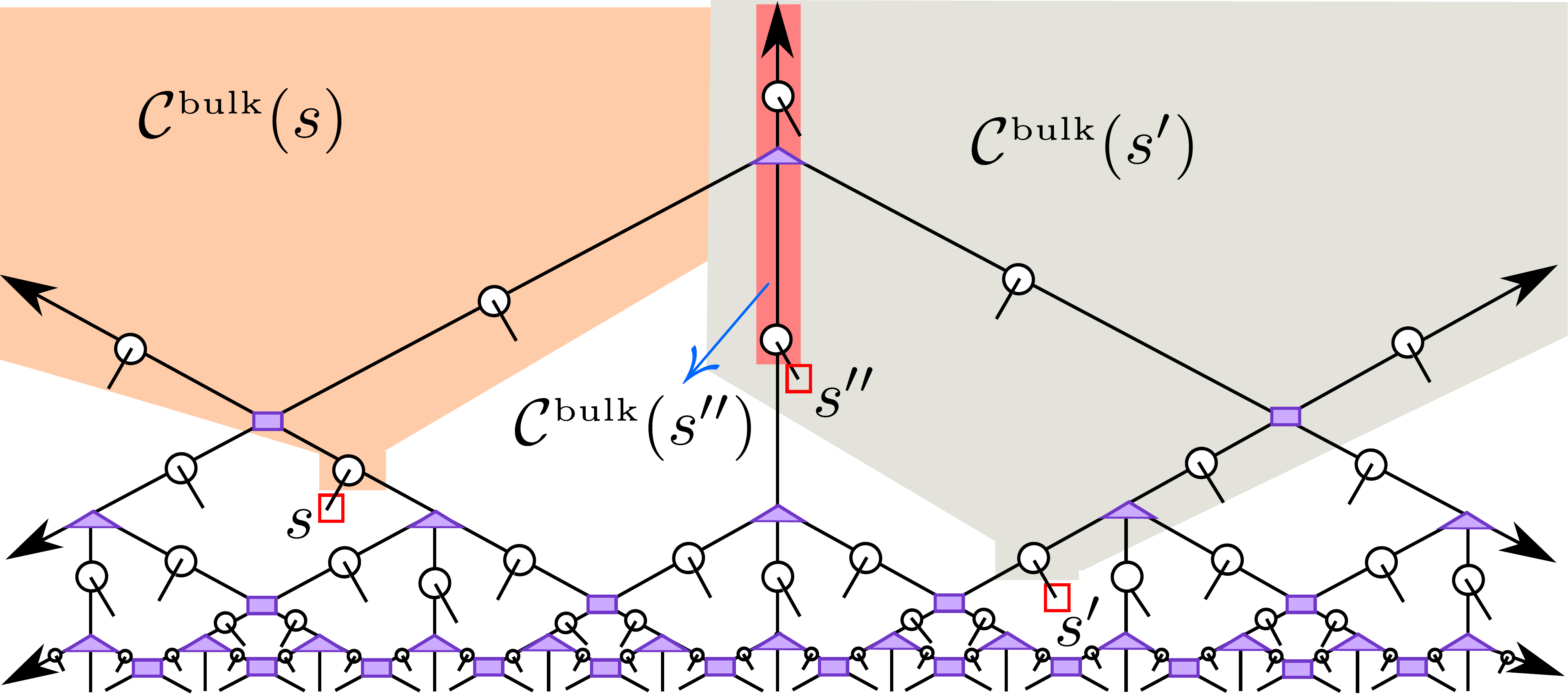}
\caption{\label{fig:causalBulk}(Color online) Causal cone structure of the lifted MERA. The reduced density matrix of a bulk site $s \in \mathcal{M}$ depends only a subset $\bulkFutureCone{s}$ of the lifted MERA tensors. $\bulkFutureCone{s}$ is the causal cone of site $s$. In particular, $\bulkFutureCone{s}$ comprises tensors which only appear at length scales larger than that of $s$, and each length scale contributes only a bounded number of tensors. Shown here are the causal cones of three different bulk sites $s,s',s'' \in \mathcal{M}$. The reduced density matrix of multiple bulk sites depends only on the tensors that belong to the union of the causal cones of the individual sites.}
\end{figure}

\begin{figure}
  \includegraphics[width=\columnwidth]{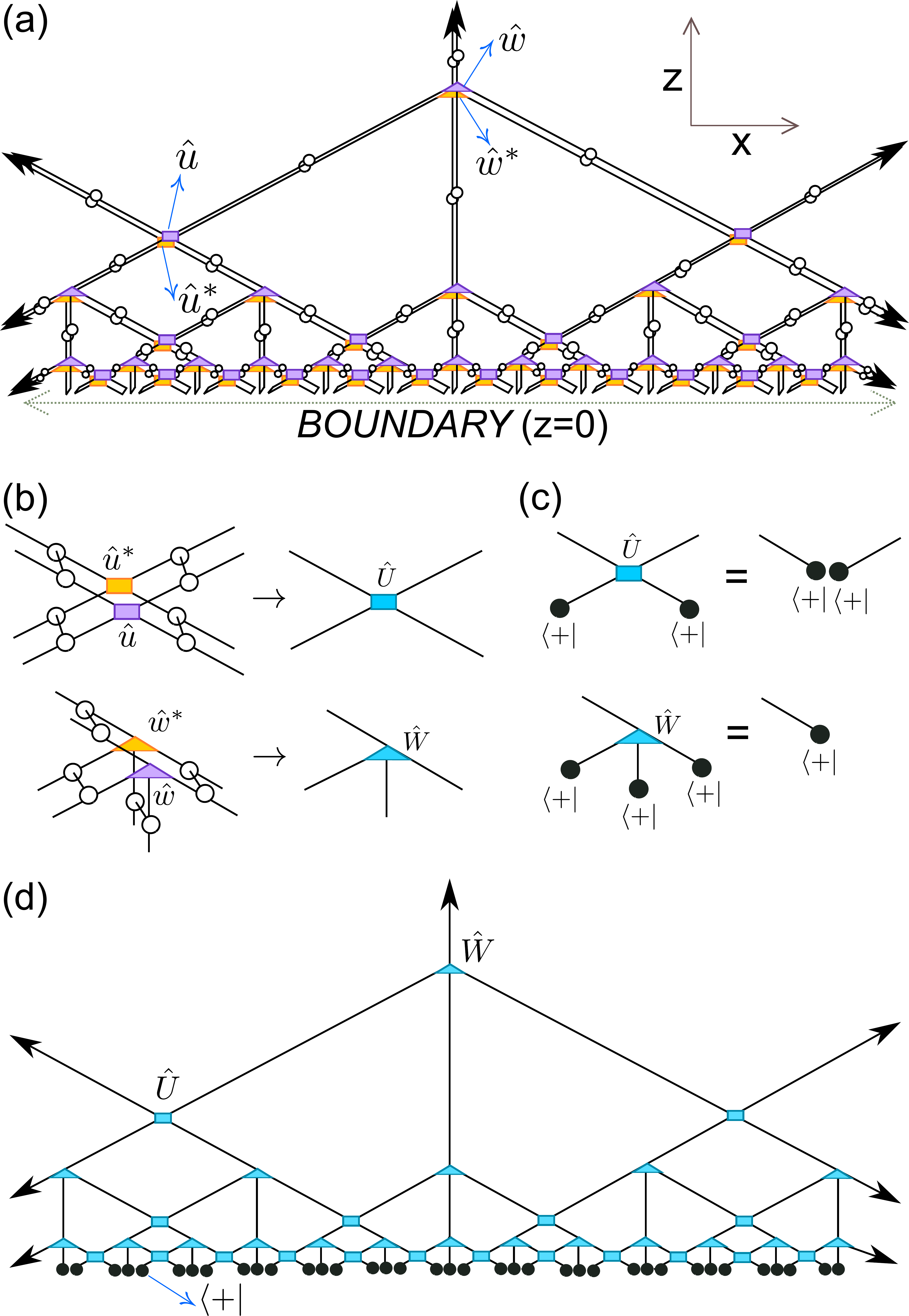}
\caption{\label{fig:bulknorm} (Color online) (a) Tensor network contraction equating to the norm of the bulk state. The open indices of the lifted MERA, including the indices located at the boundary $z=0$, are contracted with the respective open indices in the conjugate lifted MERA (obtained by replacing each tensor of the lifted MERA by its conjugate).  (b) Stochastic tensor $\hat{U}$ is obtained by contracting tensor $\hat{u}$, its complex conjugate $\hat{u}^*$, and 8 copy tensors as shown. Stochastic tensor $\hat{W}$ is obtained by contracting tensor $\hat{w}$, its complex conjugate $\hat{w}^*$, and 8 copy tensors as shown. (c) Equalities satisfied by the stochastic tensors $\hat{U}$ and $\hat{W}$. (d) The tensor network contraction shown in $(a)$ equates to a contraction of a tensor network made of copies of stochastic tensors $\hat{U},\hat{W}$ and the vector $\bra{+}$.}
\end{figure}

\begin{figure}
  \includegraphics[width=\columnwidth]{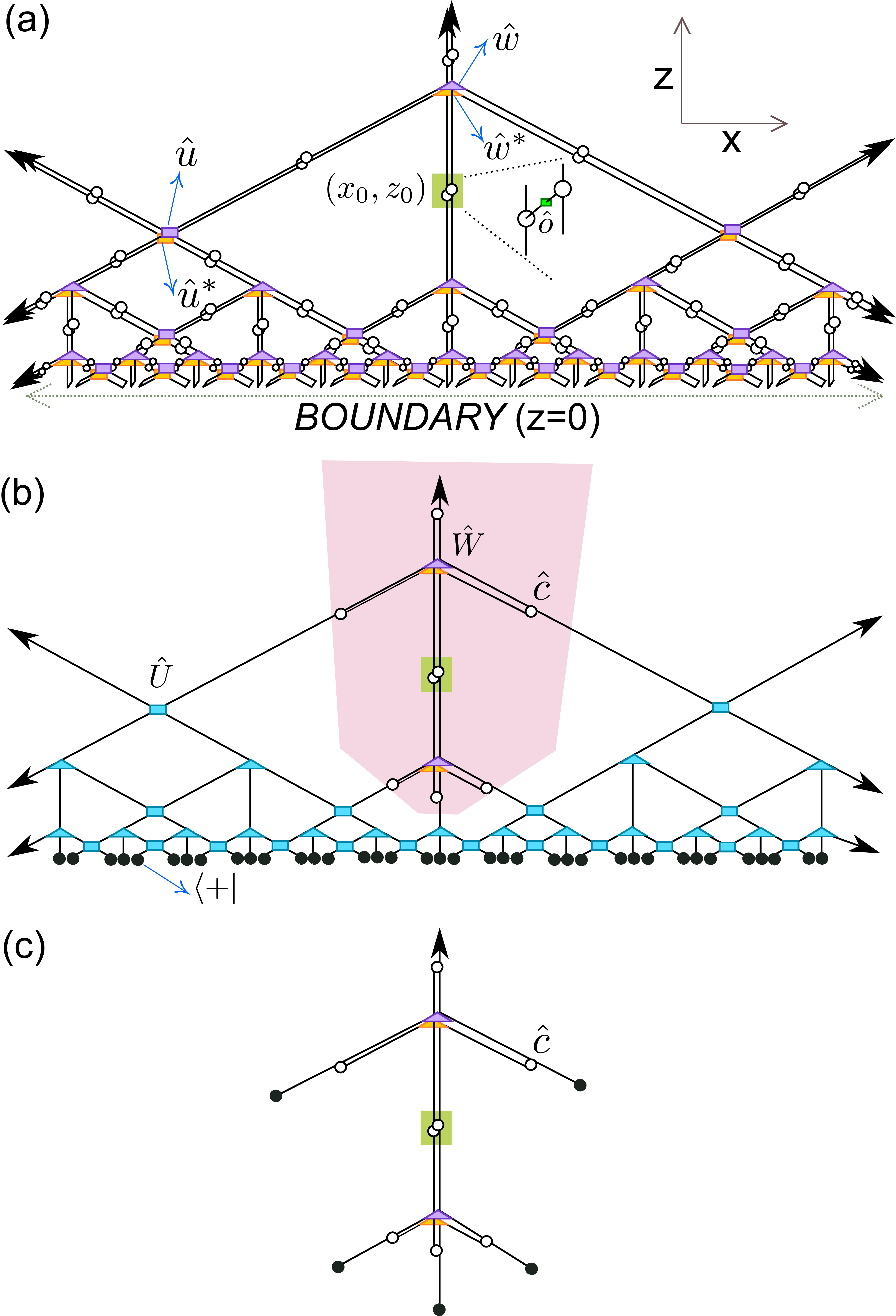}
\caption{\label{fig:onePointBulk} (Color online) (a) Tensor network contraction equating to the bulk expectation value of a one site operator $\hat{o}$ located at $(x_0,z_0)$. The contraction reduced to the contraction shown in (b), which consists of copies of the stochastic tensor $\hat{U}$ and $\hat{W}$ and some of the original MERA tensors. (c) The contraction shown in $(b)$ simplifies to a contraction of tensors only in the bulk causal cone [highlighted in $(b)$] of the site located at $(x_0,z_0)$.}
\end{figure}

Analogous to the MERA, a lifted MERA is also endowed with a bounded-width causal cone structure see \fref{fig:causalBulk}. The causal cone structure of the lifted MERA results from the fact that, in addition to tensors $\hat{u}$ and $\hat{w}$ being isometric [\eref{eq:isometric}], the copy tensor is also isometric, namely,
\begin{equation}
\sum_{qr} \hat{c}^p_{qr}(\hat{c}^*)_{p'}^{qr} = \delta^p_{p'},~~~(\hat{c}^*)_{p'}^{qr} = \hat{c}^p_{qr}.
\end{equation}
(In fact, since the copy tensor is invariant under any permutation of its indices, it is an isometry after any two of its indices are grouped into a single index.)
The causal cone structure aids in the efficient computation of expectation values from a lifted MERA.

Consider the tensor network contraction that equates to the norm of a bulk state, depicted in \fref{fig:bulknorm}(a). The contraction reduces to a contraction of a tensor network that is made of stochastic tensors and is closed with copies of the vector $\bra{+}$ at $z=0$, as shown in \fref{fig:bulknorm}(b)-(c). The contraction of the stochastic tensor network simplifies since
\begin{equation}\label{eq:stochasticprop}
\begin{split}
\hat{U} (\ket{+} \otimes \ket{+}) &= (\ket{+} \otimes \ket{+}),\\
\hat{W} (\ket{+} \otimes \ket{+} \otimes \ket{+}) &= \ket{+},
\end{split}
\end{equation}
depicted in \fref{fig:bulknorm}(d).Thus, the norm is equal to 1.

Next, consider the tensor network contraction that equates to the expectation value of an one-site operator $\hat{o}$ acting at location $(x_0,z_0)$, as illustrated in \fref{fig:onePointBulk}(a). The tensors at the boundary $z=0$ that are contracted with their Hermitian adjoints cancel out, which brings the tensors at the next length scale in contact with their Hermitian adjoints and so on. Thus, a cascade of pairwise contractions cancels all tensors located between $0 \leq z < z_0$. Furthermore, all tensors located at $z\geq z_0$ that appear outside the causal cone of the site $(x_0,z_0)$ also cancel out. Ultimately, only tensors within the causal cone of the site $(x_0,z_0)$ remain, as depicted in \fref{fig:onePointBulk}(d). The number of tensors that appear in the causal cone at increasingly large length scales is bounded and therefore the tensors within the causal cone can be contracted together efficiently and exactly by using a transfer matrix approach (a technique commonly applied in MERA algorithms \cite{MERAAlgo}). 

\section{Area law entanglement in the bulk}\label{app:arealaw}

Define the perimeter (area of the boundary) of a 2D subsystem of the bulk lattice $\mathcal{M}$ as the number of sites that are located at the subsystem's boundary. In this section, we prove that: \textit{the subsystem entanglement entropy of the bulk state described by a lifted MERA scales at most as the perimeter of the subsystem.} Such a scaling is called `(boundary) area law entanglement' in condensed matter physics, where its commonly exhibited by ground states of local 1D and 2D Hamiltonians \cite{AreaLaw}.

\begin{figure}[t]
  \includegraphics[width=\columnwidth]{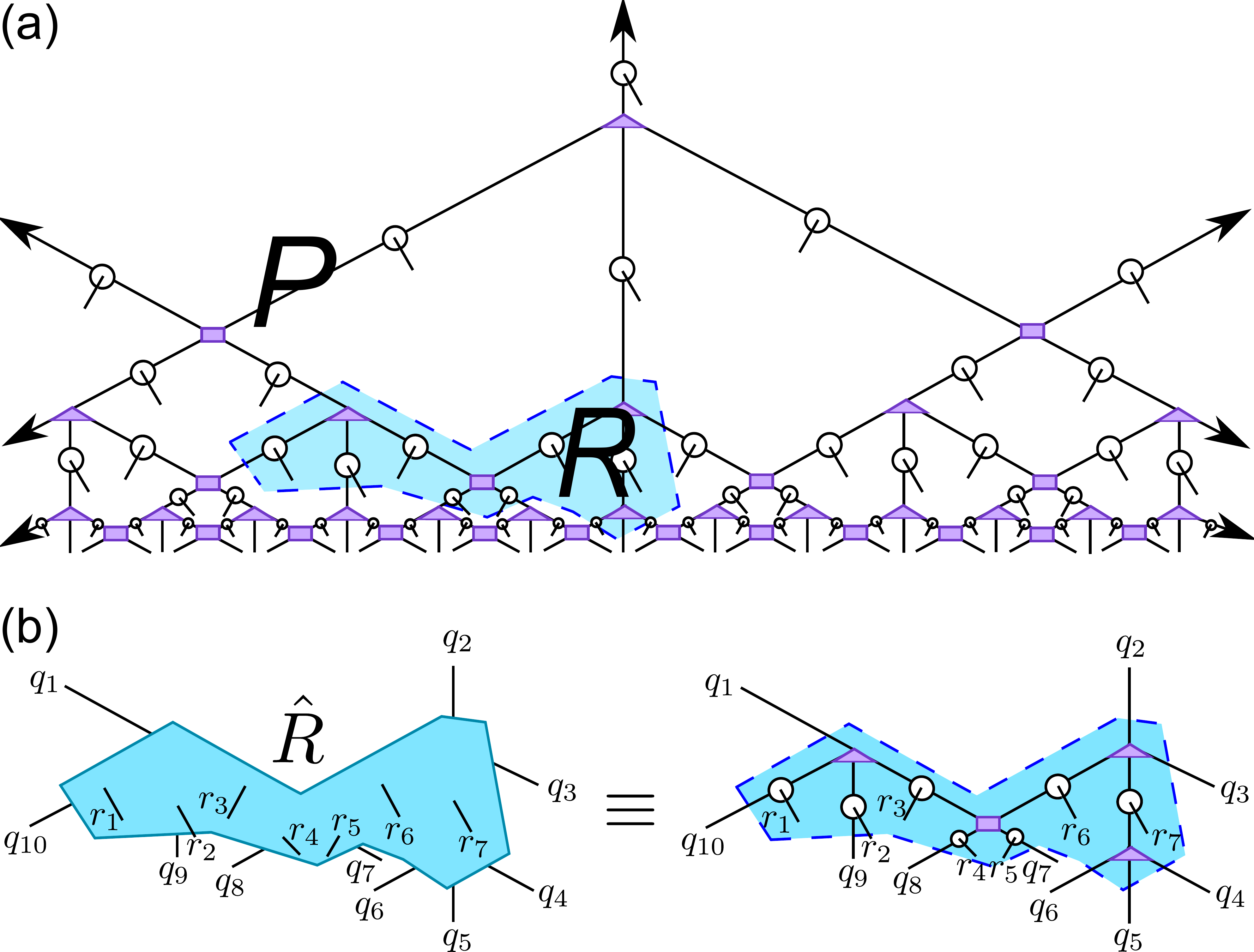}
\caption{\label{fig:bipartite} (Color online) (a) A bipartition (highlighted in blue) of the lifted MERA into a subnetworks $P$ and $R$. (b) Tensor $\hat{R}$ obtained by contracting all the tensors in $R$. Indices $q_1,q_2,\ldots,q_{10}$ connect $R$ with $P$. $r_1,r_2, \ldots, r_7$ are the open indices located inside $R$.}
\end{figure}

\textit{Proof.}---Consider a bipartition of the lifted MERA tensor network into subnetworks $P$ and $R$, see \fref{fig:bipartite}(a). This also corresponds to a bipartition of the bulk lattice $\mathcal{M}$ into parts $\mathcal{P} \subset \mathcal{M}$ and $\mathcal{R} \subset \mathcal{M}$ comprised of sites associated with the (open indices of) copy tensors in subnetworks $P$ and $R$ respectively.
The bulk state $\myphi$ can be expressed as
\begin{equation}\label{eq:bulkcollectiveindex}
\myphi = \sum_{pqr} \hat{P}_{pq} \hat{R}_{qr} \ket{p}\otimes \ket{q} \otimes \ket{r},
\end{equation}
where $p\equiv(p_1,p_2, \ldots)$, $q\equiv(q_1,q_2, \ldots)$, $r\equiv(r_1,r_2, \ldots)$, and $\hat{P}_{pq}$ and $\hat{R}_{qr}$ are obtained by contracting all tensors in subnetworks $P$ and $R$ respectively, see \fref{fig:bipartite}(b).
The reduced density matrix $\hat{\rho}^{\tiny \mathcal{R}}$ of $\mathcal{R}$ is
\begin{equation}\label{eq:rhoQ}
\begin{split}
\hat{\rho}^{\tiny\mathcal{R}} = \sum_{rr'}\sum_{q} \sum_p\hat{P}_{pq}P^*_{pq} \hat{R}_{qr}\hat{R}^*_{qr'}~\ketbra{r}{r'},
\end{split}
\end{equation}
which may be re-expressed as
\begin{equation}\label{eq:rhoMat}
\begin{split}
\hat{\rho}^{\tiny\mathcal{R}} = \sum_{rr'}(\sum_q\hat{S}_{qr} \hat{R}^*_{qr'})~\ketbra{r}{r'},
\end{split}
\end{equation}
where $\hat{S}_{qr} = \sum_{p} \hat{P}_{pq} \hat{P}^*_{pq} \hat{R}_{qr} $. We can write \eref{eq:rhoMat} in matrix form as $\hat{\rho}^{\tiny\mathcal{R}} = \hat{S}^{\tiny T}\hat{R}^*$, where matrices $\hat{S}^{\tiny T}$ and $\hat{R}^*$ have dimensions $|r| \times |q|$ and $|q| \times |r|$ respectively. Here $|i|$ denotes the number of values that index $i$ takes. This implies that if $|r| \geq |q|$ then the rank of $\hat{\rho}^{\tiny\mathcal{R}}$ is less than equal to $|q|$. This, in turn, implies that the entanglement entropy $S(\hat{\rho}^{\tiny\mathcal{R}}) = -\mbox{Tr}(\hat{\rho}^{\tiny\mathcal{R}} \mbox{log}_2 \hat{\rho}^{\tiny\mathcal{R}})$ of a sufficiently large subsystem $\mathcal{R}$ is O(log~$|q|$), where log$~|q|$ is the number of sites that comprise the boundary of $\mathcal{R}$. Thus, the bulk subsystem entanglement entropy scales at most as the perimeter of the subsystem.$\blacksquare$

Note that the proof does not depend on the components of the tensors, but only on the fact that the bulk state is encoded in a lifted MERA tensor network. We remark that (boundary) area law entanglement scaling proved above is subtle in a 2D hyperbolic geometry (here the hyperbolic lattice $\mathcal{M}$) where the area of a subsystem can be proportional to its perimeter.


\section{Constrained MERA representations}\label{app:unitaryFreedom}
In this appendix, we describe how to translate the standard MERA representation, reviewed in Sec.~\ref{sec:boundary}, to a MERA representation in which the intrinsic bond freedom is significantly constrained. See discussion in Sec.~\ref{sec:bulkentanglement}. The plots in \fref{fig:results1} and \fref{fig:results2} were obtained by using the constrained MERA representation described next.

\subsection{Higher order singular value decomposition}
Define the norm of a tensor $\hat{T}_{i_1i_2\ldots i_n}$ as
\begin{equation}
||\hat{T}|| = \sum_{i_1i_2\ldots i_n} \hat{T}_{i_1i_2\ldots i_n}\hat{T}^{*}_{i_1 i_2 \ldots i_n}.
\end{equation}
Any tensor $\hat{T}_{i_1i_2\ldots i_n}$ can be decomposed into a `core tensor' $\hat{S}_{i_1i_2\ldots i_n}$ and a set of $n$ unitary matrices $\hat{U}^{[k]}$ ($k\in \{1,2,\ldots,n\}$), one for each index of $\hat{T}$ such that
\begin{equation}\label{eq:hosvd}
\hat{T}_{i_1i_2\ldots i_n} = \sum_{i_1 i_2 \ldots i_n} \hat{S}_{i'_1 i'_2 \ldots i'_n}(\hat{U}^{[1]})^{i'_1}_{i_1}(\hat{U}^{[2]})^{i'_2}_{i_2}\cdots(\hat{U}^{[n]})^{i'_n}_{i_n},
\end{equation}
and the core tensor $\hat{S}$ satisfies
\begin{equation}\label{eq:core}
||\hat{S}^{[m,p]}|| \geq ||\hat{S}^{[m,p']}||,~~~\mbox{if }p > p', ~~\forall m.
\end{equation}
Here $\hat{S}^{[m,p]}$ is a $(n-1)$-index tensor obtained from $\hat{S}$ by fixing the $m^{th}$ index of $\hat{S}$ to value $p$, that is,
\begin{equation}
(\hat{S}^{[m,p]})_{j_1 j_2 j_{n-1}} \equiv \hat{S}_{j_1 \ldots j_m=p \ldots j_{n-1}}.
\end{equation} 
Equation \ref{eq:hosvd}, along with the constraint \eref{eq:core}, is known as the \textit{higher order singular value decomposition} (HOSVD). It can be obtained by means of a sequence of matrix singular value decompositions \cite{HOSVD}. Note that if tensor $\hat{T}$ can be reshaped into an isometric matrix according to some bipartition of its indices then the core tensor $\hat{S}$ is also an isometry according to the same bipartition.
Also, if $\hat{T}$ is a unitary tensor (according to some bipartition of its indices) it can be easily shown that $||\hat{T}^{[m,p]}||$ is constant for all $m,p$. This means that given an HOSVD decomposition of a unitary tensor one can contract the core tensor $\hat{S}$ with an \textit{arbitrary} unitary matrix $\hat{R}$ on any index $k$ and update $\hat{U}^{[k]'} = \hat{R}^\dagger \hat{U}^{[k]}$ to obtain another HOSVD of the tensor. On the other hand, the most constrained HOSVD possible for a generic tensor is unique up to contractions with only \textit{diagonal} unitary matrices.

\begin{figure}
  \includegraphics[width=\columnwidth]{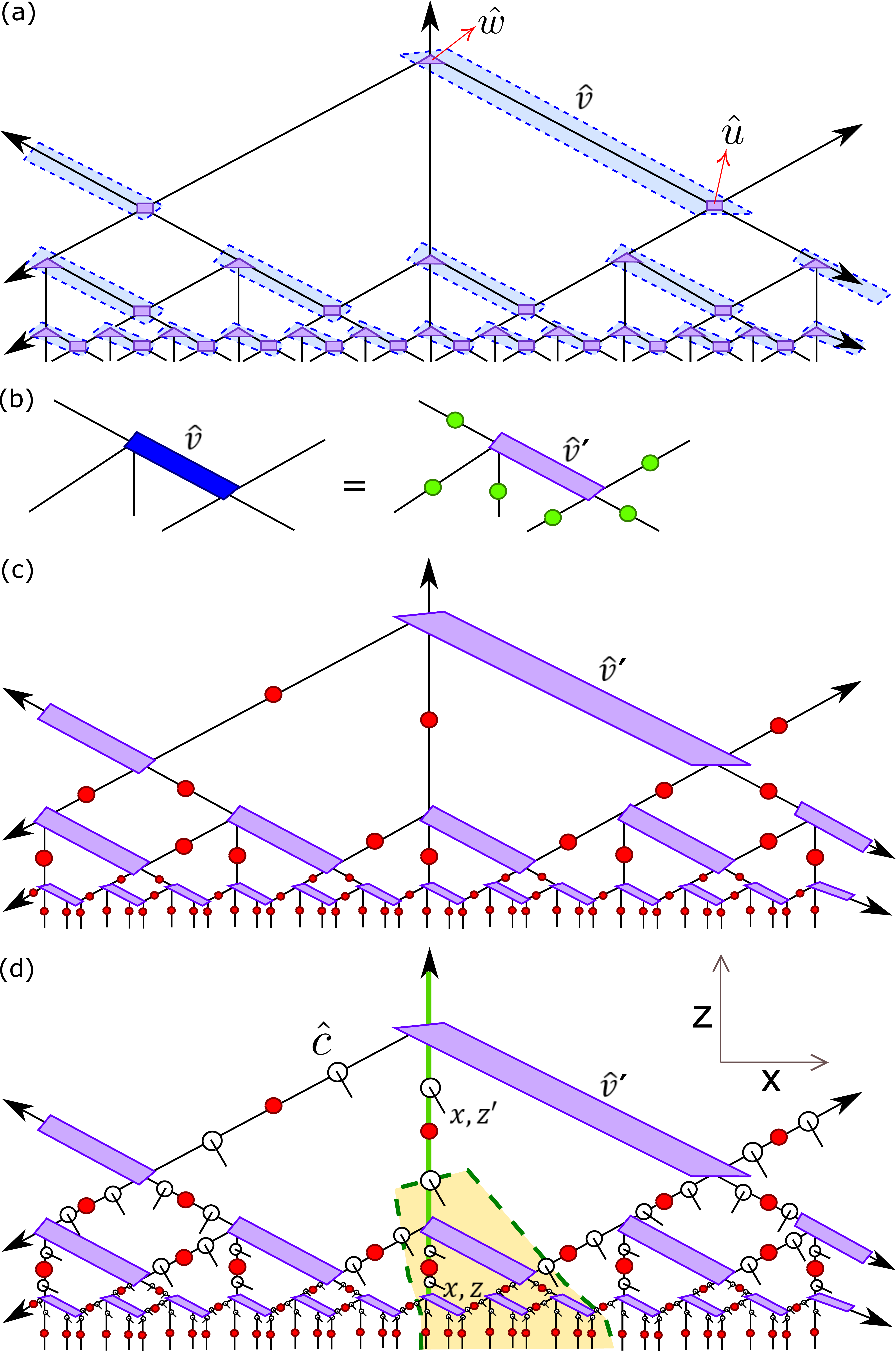}
\caption{\label{fig:canonical} (Color online) (a) Coarse-graining the MERA by contracting pairs of tensors $\hat{u}$ and $\hat{w}$ to obtain an isometric tensor $\hat{v}$. (b) Higher order singular value decomposition, \eref{eq:hosvd}, of the coarse-grained isometric tensor $\hat{v}$ into an isometric core tensor $\hat{v}'$ (purple) and a unitary matrix (yellow) on each bond. (c) HOSVD-MERA obtained by decomposing each $\hat{v}$ as shown in $(b)$ and multiplying the two green unitary matrices that appear on each bond to obtain another unitary matrix (red). (d) Lifting the HOSVD-MERA by inserting copy tensors on all its bond indices. \fref{fig:results1} plots the Reyni entanglement entropy of bulk sites located on a geodesic holographic screen, illustrated here as the green dashed path. \fref{fig:results2} plots the mutual information between two bulk sites located at different length scales e.g. the bulk sites located at $(x,z)$ and $(x,z')$ along the vertical geodesic (solid green).}
\end{figure}

Our goal here is to introduce a MERA representation that has a more constrained bond freedom than the standard MERA representation. Let us define a MERA representation by requiring that each tensor of the MERA
\begin{enumerate}
\item  is unitary or isometric, and 
\item  satisfies \eref{eq:core}.
\end{enumerate}
We refer to this MERA representation as an \textit{HOSVD-MERA}. Given a MERA representation, decompose each tensor according to \eref{eq:hosvd} and subsequently multiply together the two unitary matrices that appear on each bond. The resulting tensor network, an HOSVD-MERA, is comprised of core tensors (which replace the original tensors) and a set of unitary matrices, one for each bond.

However, as mentioned previously, if the given MERA contains unitary tensors the corresponding HOSVD-MERA representation is not any more constrained. On the other hand, HOSVD is likely to be more constrained for a tensor that is isometric (according to some bipartition of indices) but is not unitary according to \textit{any} bipartition of indices. That is, the contraction of the core tensor in the HOSVD of a generic isometric tensor with an arbitrary unitary matrix $\hat{R}$ generally breaks the constraint \eref{eq:core}.
We can exploit this fact by contracting pairs of MERA tensors, $\hat{w}$ and $\hat{u}$, to obtain a coarse-grained MERA made of only isometric tensors as depicted in \fref{fig:canonical}(a).
We then translate the coarse-grained MERA to an HOSVD-MERA representation as described above, namely, by applying HOSVD for each coarse-grained tensor, see \fref{fig:canonical}(b)-(c).

The bond freedom in an HOSVD-MERA representation of a quantum state is more constrained than the standard MERA representation since its tensors are unique up to contractions only with diagonal unitary matrices. Subsequently, the different bulk states obtained by lifting the HOSVD-MERA, after applying different \textit{diagonal} unitary bond transformations, are related to one another by the action of one-site diagonal unitary transformations on the bulk lattice. This is because the copy tensor commutes with a diagonal matrix, namely, a contraction of the copy with a diagonal matrix on any index is equal to a contraction of the copy with the same diagonal matrix on a different index. Thus, in this case, the different bulk states, dual to the same boundary state, (at least) have the same entanglement.

\subsection{A matrix form of the MERA}
\begin{figure}
  \includegraphics[width=\columnwidth]{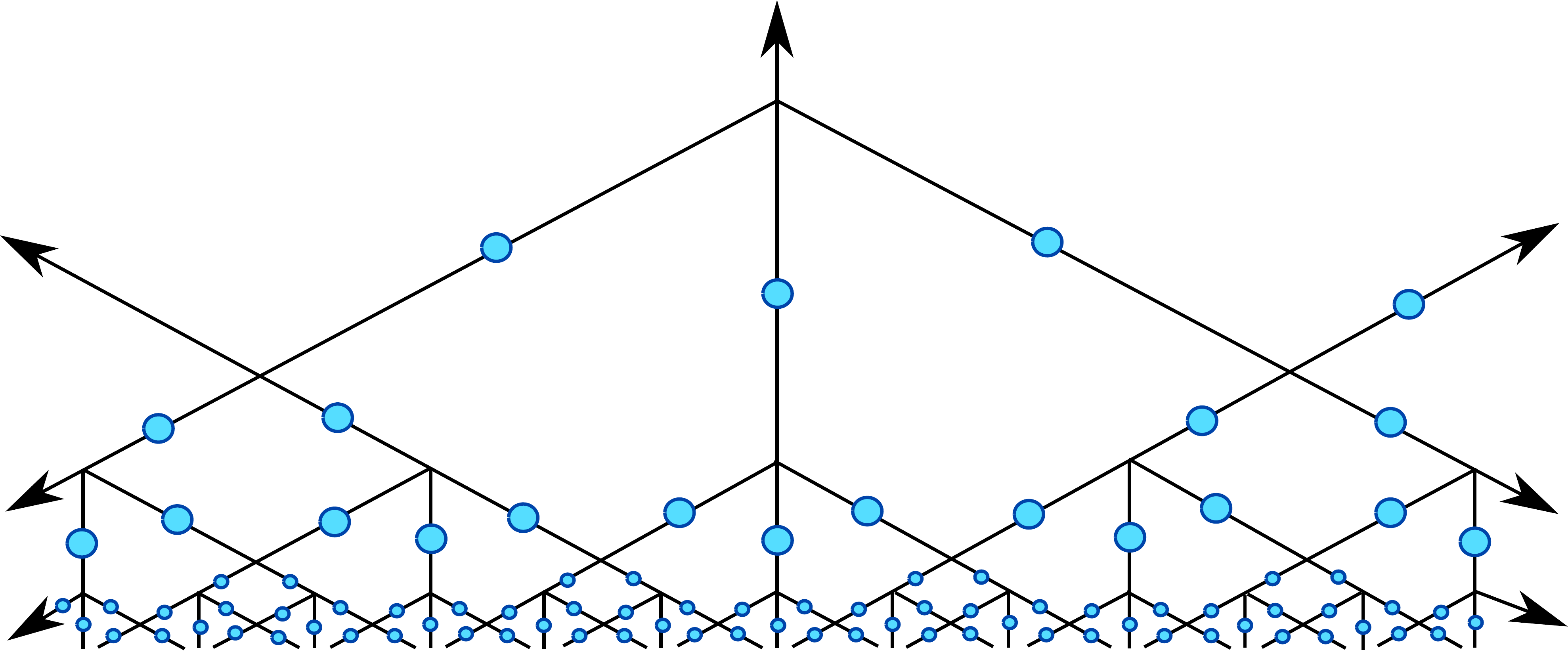}
\caption{\label{fig:matrixForm} (Color online) A matrix form of the MERA obtained by applying tensor rank decomposition for each tensor of the MERA. All the information of the quantum state represented by the MERA is collected into a set of matrices (blue), one for each bond of the MERA. Each vertex of the tensor network depicts a 4-index copy tensor that connects four matrices.}
\end{figure}

Another constrained MERA representation can be obtained by means of the so called \textit{tensor rank decomposition} of the MERA tensors. Tensor rank decomposition is an HOSVD where the core tensor is fixed to be the copy tensor \cite{TensorRank}. (The components of an $n$-index copy tensor are 1 for equal values of the $n$ indices and 0 otherwise.) For a generic tensor, tensor rank decomposition is generally unique up to systematic permutations of the bond matrices. Given a MERA representation, let us apply tensor rank decomposition for each tensor of the MERA and subsequently multiply together the two matrices that appear on each bond, see \fref{fig:matrixForm}. In the resulting tensor network representation, all the information of the quantum state is captured in the bond matrices (since all remaining tensors are fixed as copy tensors). We refer to this representation as the \textit{matrix form} of the MERA.

Note that a given MERA cannot be transformed into the matrix form by inserting resolutions of identity on the bonds, that is, by choosing a particular basis for the tensors. If the original MERA has bond dimension $\chi$, the matrix form may have a bond dimension up to $\chi' = \chi^3$. We remark that finding the \textit{optimal} tensor rank decomposition (corresponding to the smallest possible $\chi'$) of a generic tensor is NP-hard \cite{TensorRank}. However, several algorithms are available to determine a non-optimal tensor rank decomposition of a generic tensor.


\section{Proof of \eref{eq:holoScreen} for holographic screens}\label{app:holo}
\begin{figure}[t]
  \includegraphics[width=\columnwidth]{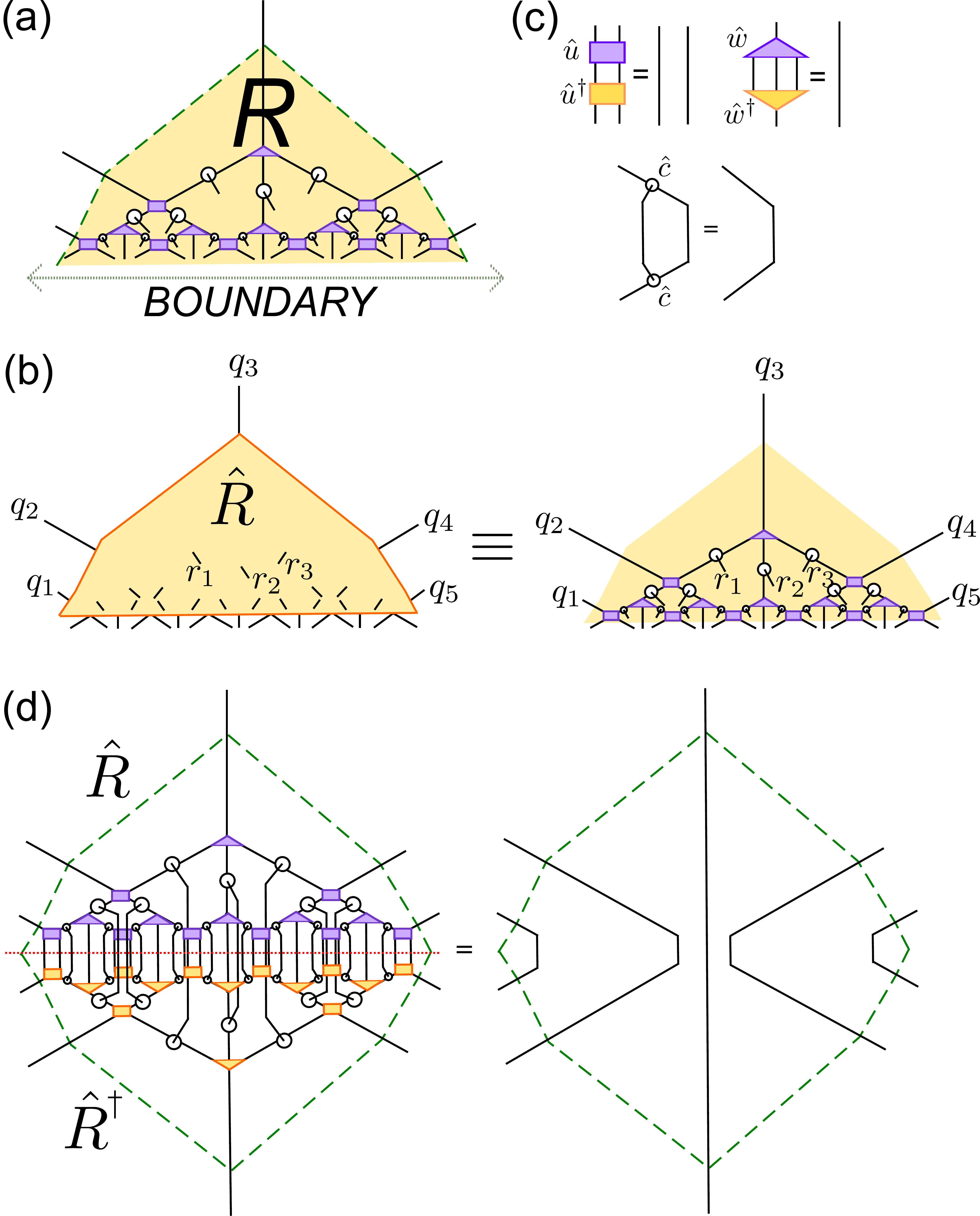}
\caption{\label{fig:holoEntProof}(Color online) (a) A separate view of the interior $R$ of the holographic screen depicted in \fref{fig:holoEnt}. (b) Tensor $\hat{R}$ that is obtained by contracting all tensors in the interior of the holographic screen, \eref{eq:ttemp}.  (c) Equalities satisfied by the isometric tensors $\hat{u}$, $\hat{w}$ and the copy tensor $\hat{c}$.  (d) Contraction of $\hat{R}$ and its adjoint $\hat{R}^\dagger$ is equal to the identity, \eref{eq:holocond}. (Thus, $\hat{R}$ is an isometry.) This equality is obtained by applying the equalities depicted in $(b)$ to pairwise contractions of tensors on the left hand side.}
\end{figure}

Let $R$ denote the set of tensors located in the interior of a holographic screen, see \fref{fig:holoEntProof}(a). Denote by $\hat{R}$ the tensor obtained by contracting together all tensors in $R$. Using components we write
\begin{equation}\label{eq:ttemp}
\hat{R} \equiv \sum_{qr} \hat{R}_{qr} \ket{q}\bra{r},
\end{equation}
where $r$ is the tuple of all open indices in the interior and $q$ is the tuple of all bond indices that connect the interior with the remaining tensor network, see \fref{fig:holoEntProof}(b). 
Because the holographic screen satisfies the arrow criterion stated in Sec.~\ref{sec:bulkholo}, it does not simultaneously intersect both an incoming and an outgoing index of any tensor of the lifted MERA. This means that tensor $\hat{R}$ is simply a composition of copies of the isometric tensors $\hat{u},\hat{w},$ and $\hat{c}$ [\fref{fig:holoEntProof}(c)], and therefore $\hat{R}$ itself is also an isometry satisfying
\begin{equation}\label{eq:holocond}
\hat{R}\hat{R}^\dagger = \hat{I},~~~~~~\sum_r \hat{R}_{qr} \hat{R}^*_{q'r} = \delta_{qq'},
\end{equation}
as depicted in \fref{fig:holoEntProof}(d).

Let us expand the bulk state according to \eref{eq:bulkcollectiveindex}. The reduced density matrix $\hat{\rho}^{\tiny \mbox{screen}}_{\tiny \mbox{geo}}$ of the bulk sites located on the screen is
\begin{equation}\label{eq:rhoB0}
\hat{\rho}^{\tiny \mbox{screen}}_{\tiny \mbox{geo}} =\sum_{qq'}(\sum_{p} \hat{P}_{pq}\hat{P}^*_{pq'})(\sum_r \hat{R}_{qr} \hat{R}^*_{q'r})\ketbra{q}{q'}, 
\end{equation}
where $p$ is the tuple of all open indices in the exterior and tensor $\hat{P}$ is obtained by contracting together all tensors in the exterior subnetwork. Using \eref{eq:holocond}, \eref{eq:rhoB0} simplifies to
\begin{equation}\label{eq:rhoB1}
\hat{\rho}^{\tiny \mbox{screen}}_{\tiny \mbox{geo}} =\sum_{qq'}(\sum_{p} \hat{P}_{pq}\hat{P}^*_{pq'})\delta_{qq'}\ketbra{q}{q'} = \sum_{qp} \hat{P}_{pq}\hat{P}^*_{pq}\ketbra{q}{q}.
\end{equation}

Next, the reduced density matrix $\hat{\rho}^{\tiny \mbox{interior}}_{\tiny \mbox{geo}}$ of all the bulk sites in the interior is
\begin{equation}\label{eq:t0}
\hat{\rho}^{\tiny \mbox{interior}}_{\tiny \mbox{geo}} = \sum_{rr'}\sum_{q}\sum_{p} \hat{P}_{pq}\hat{P}^*_{pq}\hat{R}_{qr}\hat{R}^*_{qr'}) \ket{r}\bra{r'},
\end{equation}
which we re-arrange slightly as
\begin{equation}\label{eq:t1}
\hat{\rho}^{\tiny \mbox{interior}}_{\tiny \mbox{geo}} = \sum_{q}(\sum_{p} \hat{P}_{pq}\hat{P}^*_{pq})(\sum_{r}\hat{R}_{qr} \ket{r}) (\sum_{r'}\hat{R}^*_{qr'}\bra{r'}).
\end{equation}
Multiplying $\ket{r}$ on both sides of \eref{eq:ttemp} we obtain
\begin{equation}\label{eq:ttemp1}
\hat{R}\ket{r} = \sum_{q} \hat{R}_{qr} \ket{q}.
\end{equation}
Using \eref{eq:ttemp1} and \eref{eq:rhoB1} in \eref{eq:t1} we have
\begin{equation}
\begin{split}
\hat{\rho}^{\tiny \mbox{interior}}_{\tiny \mbox{geo}} &=\hat{R}^\dagger (\sum_{qp} \hat{P}_{pq}\hat{P}^*_{pq}\ketbra{q}{q})\hat{R}\\
&=\hat{R}^\dagger (\hat{\rho}^{\tiny \mbox{screen}}_{\tiny \mbox{geo}})\hat{R},
\end{split}
\end{equation}
which proves \eref{eq:holoScreen}.$\blacksquare$

Note that the above proof does not rely on the actual components of the tensors. Thus, the presence of holographic screens is a general structural property of a copy lifted MERA.

\begin{figure*}
  \includegraphics[width=18cm]{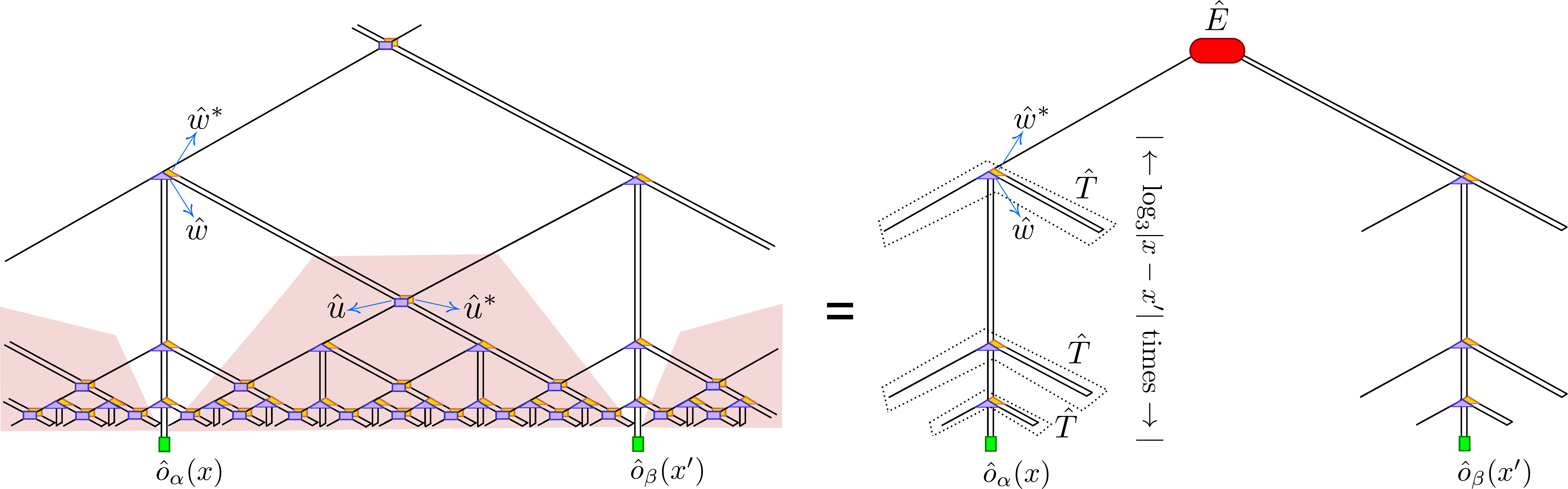}
\caption{\label{fig:corr} (Color online) (Left) Tensor network contraction equating to the two point correlator $\xbound{\hat{o}_{\alpha}(x) \hat{o}_{\beta}(x')}$ of one-site operators $\hat{o}_{\alpha}$ and $\hat{o}_{\beta}$. The contraction consists of gluing the open indices of the MERA with the corresponding open indices of the conjugate MERA (namely, the tensor network obtained by replacing each tensor of MERA with its complex conjugate, shown in yellow), and sandwiching operators $\hat{o}_{\alpha}$ and $\hat{o}_{\beta}$ on the open indices located at $x$ and $x'$ respectively. Tensors located inside the  highlighted red regions are pairwise mutiplied with their Hermitian adjoints and cancel out. The resulting contraction (shown on the right) consists of $p \equiv \mbox{log}_3(|x-x'|)$ powers of a transfer superoperator $\hat{T}$ (obtained by contracting tensors $\hat{w}$ and $\hat{w}^*$ enclosed within a dotted contour) and the environment tensor $\hat{E}$ that is obtained by contracting all remaining tensors  appearing at length scales larger than $p$.}
\end{figure*}

\begin{figure*}
  \includegraphics[width=18cm]{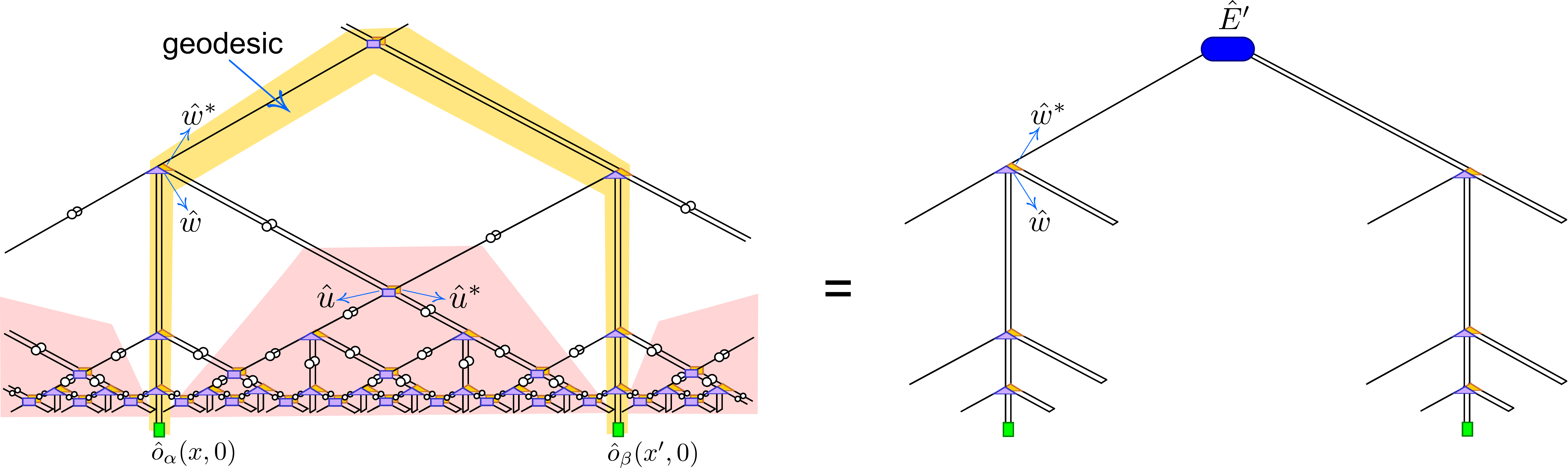}
\caption{\label{fig:bulkcorr} (Color online) (Left) Tensor network contraction equating to the bulk expectation value of the string operator $\hat{K}_{x,x'}\hat{o}_\alpha({x,0}) \hat{o}_\beta({x',0})$. The contraction consists of gluing the open indices of the lifted MERA with the corresponding open indices of the conjugate lifted MERA, sandwiching operators $\hat{o}_{\alpha}$ and $\hat{o}_{\beta}$ on the open indices located at $x$ and $x'$ respectively, and sandwiching the operator $\hat{K}_{x,x'}$ on the open indices located along the graph geodesic (highlighted yellow) extending between $(x,0)$ and $(x',0)$. $\hat{K}_{x,x'}$ acts to remove the copy tensors on the geodesic. (Note that the copy tensors inside the yellow region have been removed.) Tensors located inside the highlighted red regions are pairwise multiplied with their Hermitian adjoints and cancel out. The resulting contraction (shown on the right) consists of $p \equiv \mbox{log}_3(|x-x'|)$ powers of the same transfer superoperator $\hat{T}$ that appears in \fref{fig:corr} but a different environment tensor $\hat{E}'$, which is obtained by contracting all remaining tensors of the lifted MERA appearing at length scales larger than $p$.}
\end{figure*}

\section{Proof of the bulk/boundary dictionary} \label{app:critical}

In this appendix, we prove the formulae Eqs.~(\ref{eq:corr})-(\ref{eq:reyni}) that were listed in Sec.~\ref{sec:dictionary} of the main text.
The proofs essentially employ two properties: (i) the causal cone structure of the lifted MERA (illustrated in \fref{fig:causalBulk}), and (ii) the fact that the boundary state can be recovered from a bulk state by projecting each bulk site to the state $\ket{+} = \sum_{j=1}^\chi \ket{j}$, see \eref{eq:recoverstate}.  We prove each formula by illustrating that the tensor network contractions equating to the left hand side and the right hand side of the formula, are equal (up to a multiplicative or additive constant).
[\textit{Note.--} The notation and symbols used in this Appendix are introduced in Sec.~\ref{sec:dictionary}.]

\begin{figure*}
  \includegraphics[width=18cm]{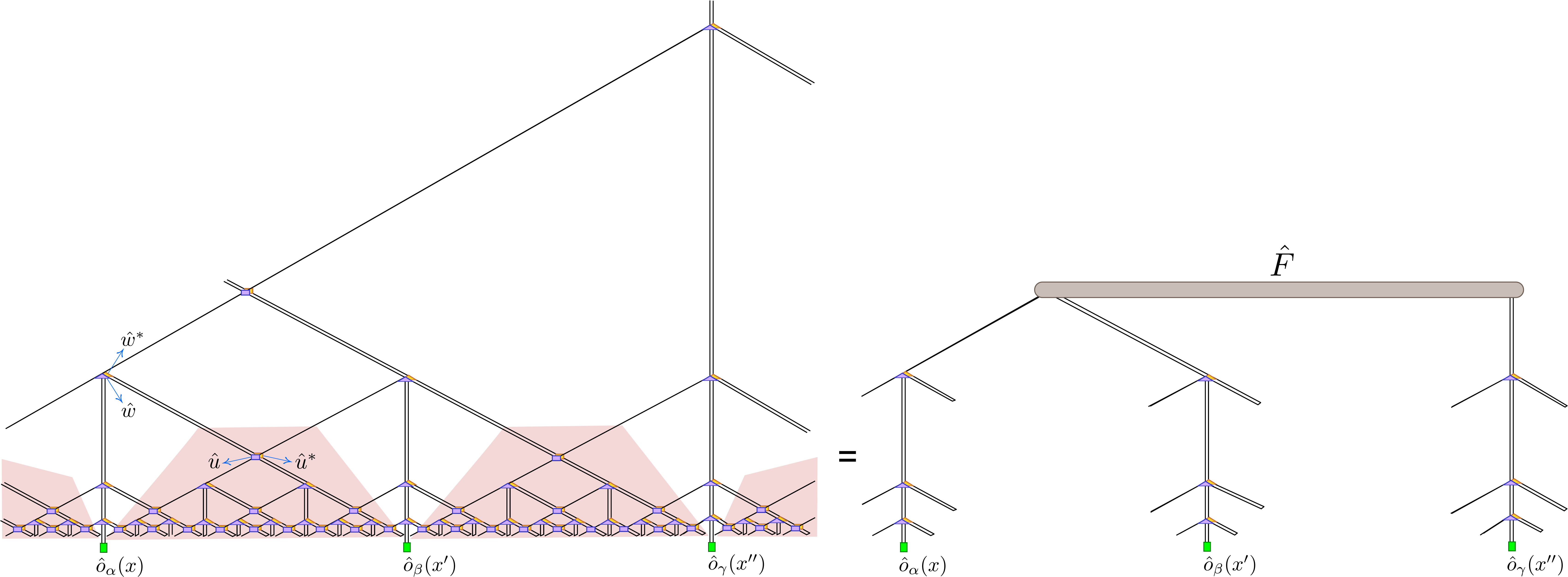}
\caption{\label{fig:ope}  (Color online) (Left) Tensor network contraction equating to the three point correlator $\xbound{\hat{o}_{\alpha}(x) \hat{o}_{\beta}(x')\hat{o}_{\beta}(x'')}$ where e.g. $|x-x'| = |x'-x''| = \ell = 27$. The contraction consists of gluing the open indices of the MERA with the corresponding open indices of the conjugate MERA, and sandwiching operators $\hat{o}_{\alpha}, \hat{o}_{\beta}$ and $\hat{o}_{\gamma}$ on the open indices located at $x,x'$ and $x''$ respectively. Tensors located inside the highlighted red regions are pairwise mutiplied with their Hermitian adjoints and cancel out. The resulting contraction (shown on the right) consists of $\mbox{log}_3\ell$ powers of the same transfer superoperator $\hat{T}$ that appears in \fref{fig:corr} and an environment tensor $\hat{F}$, which is obtained by contracting all remaining tensors appearing at length scales larger than $\mbox{log}_3\ell$.}
\end{figure*}

\begin{figure*}
  \includegraphics[width=18cm]{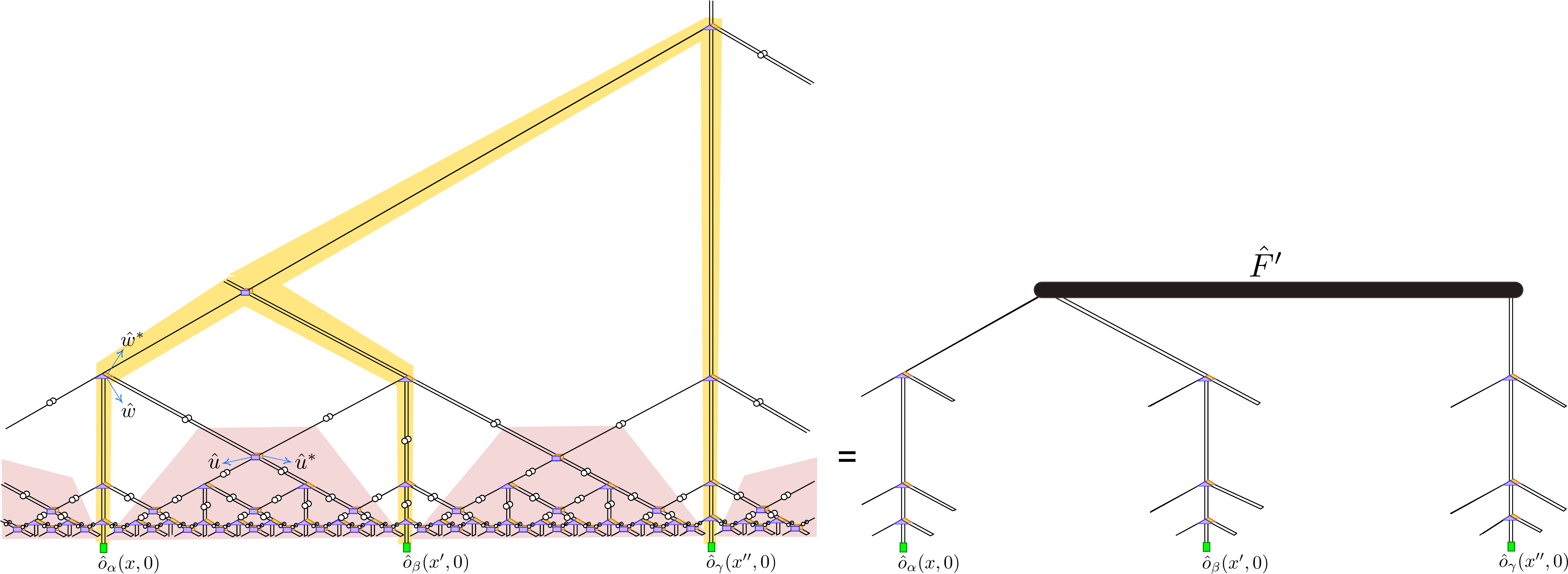}
\caption{\label{fig:bulkope}  (Color online) (Left) Tensor network contraction equating to the bulk expectation value of the extended operator $\hat{T}_{x,x',x''}\hat{o}_\alpha({x,0}) \hat{o}_\beta({x',0})  \hat{o}_\gamma({x'',0})$  where e.g. $|x-x'| = |x'-x''| = \ell = 27$. The contraction consists of gluing the open indices of the lifted MERA with the corresponding open indices of the conjugate lifted MERA, sandwiching operators $\hat{o}_{\alpha}$ and $\hat{o}_{\beta}$ on the open indices located at $x$ and $x'$ respectively, and sandwiching the operator $\hat{T}_{x,x',x''}$ on the open indices indices located on the union (highlighted yellow) of the graph geodesic between $x$ and $x'$ and the graph geodesic between $x$ and $x''$ respectively. $\hat{T}_{x,x',x''}$ acts to remove the copy tensors within the highlighted yellow region. Tensors located inside the red highlighted regions are pairwise mutiplied with their Hermitian adjoints and cancel out. The resulting contraction (shown on the right) consists of $\mbox{log}_3\ell$ powers of the same transfer superoperator $\hat{T}$ that appears in \fref{fig:ope} but a different  is environment tensor $\hat{F}'$, which is obtained by contracting all remaining tensors appearing at length scales larger than $\mbox{log}_3\ell$.}
\end{figure*}

\subsection{2-point correlators}
The 2-point boundary correlator $\xbound{\hat{o}_\alpha(x) \hat{o}_\beta({x'})}$ of scaling operators $\hat{o}_\alpha$ and $\hat{o}_\beta$ acting at locations $x$ and $x'$, $|x-x'| = 3^q$ ($q$ is a positive integer), is obtained by a tensor network contraction illustrated on the left hand side in \fref{fig:corr}.
Since tensors $\hat{u}$ and $\hat{w}$ are isometric, all tensors that are multiplied with their Hermitian adjoints cancel and the contraction simplifies to the one shown on the right hand side in \fref{fig:corr}. Here $\hat{E}$ is the environment tensor obtained by contracting together all tensors at length scales above $q$ that do not cancel out (that is, tensors located inside the causal cone).
Using the fact that operators $\hat{o}_\alpha$ and $\hat{o}_\beta$ are eigenoperators of the scaling superoperator $\hat{T}$ with eigenvalues $\lambda_\alpha$ and $\lambda_\beta$ respectively, we obtain the closed expression
\begin{equation}\label{eq:corr2bound}
\xbound{\hat{o}_\alpha(x) \hat{o}_\beta({x'})} = \frac{C_{\alpha\beta}}{|x-x'|^{\Delta_{\alpha} + \Delta_{\beta}}},
\end{equation}
where $C_{\alpha\beta} = \mbox{Tr}[\hat{E}(\hat{o}_\alpha \otimes \hat{o}_\beta)]$, and $\Delta_{\alpha} = -\mbox{log}_3 \lambda_\alpha$ and $\Delta_{\beta} = -\mbox{log}_3 \lambda_\beta$ are the scaling dimensions of the scaling operators $\hat{o}_\alpha$ and $\hat{o}_\beta$ respectively.

Next, the corresponding bulk expectation value $\xbulk{\hat{K}_{x,x'}\hat{o}_\alpha({x,0}) \hat{o}_\beta({x',0})}$ is obtained by contracting the tensor network depicted on the left hand side in \fref{fig:bulkcorr}. Using the fact that tensors $\hat{u}$ and $\hat{w}$ are isometric, the contraction simplifies to the one shown on the right hand side in \fref{fig:bulkcorr}, and we obtain the expression
\begin{equation}\label{eq:corr2bulk}
\xbulk{\hat{K}_{x,x'}\hat{o}_\alpha({x,0}) \hat{o}_\beta({x',0})} = 
\frac{C'_{\alpha\beta}}{|x-x'|^{\Delta_{\alpha} + \Delta_{}\beta}},
\end{equation}
where $C'_{\alpha\beta} = \mbox{Tr}[\hat{E'}(\hat{o}_\alpha \otimes \hat{o}_\beta)]$.
Comparing \eref{eq:corr2bound} and \eref{eq:corr2bulk}, we have
\begin{equation}
\begin{split}
\xbound{\hat{o}_\alpha(x) \hat{o}_\beta({x'})} = f(\hat{u},\hat{w},\hat{o}_{\alpha},\hat{o}_{\beta}) \times & \\
\xbulk{\hat{K}_{x,x'}\hat{o}_\alpha({x,0}) \hat{o}_\beta({x',0})}&,
\end{split}
\end{equation}
where $f(\hat{u},\hat{w},\hat{o}_{\alpha},\hat{o}_{\beta}) = C_{\alpha\beta}/C'_{\alpha\beta}$, which is \eref{eq:corr}.

\begin{figure*}
  \includegraphics[width=18cm]{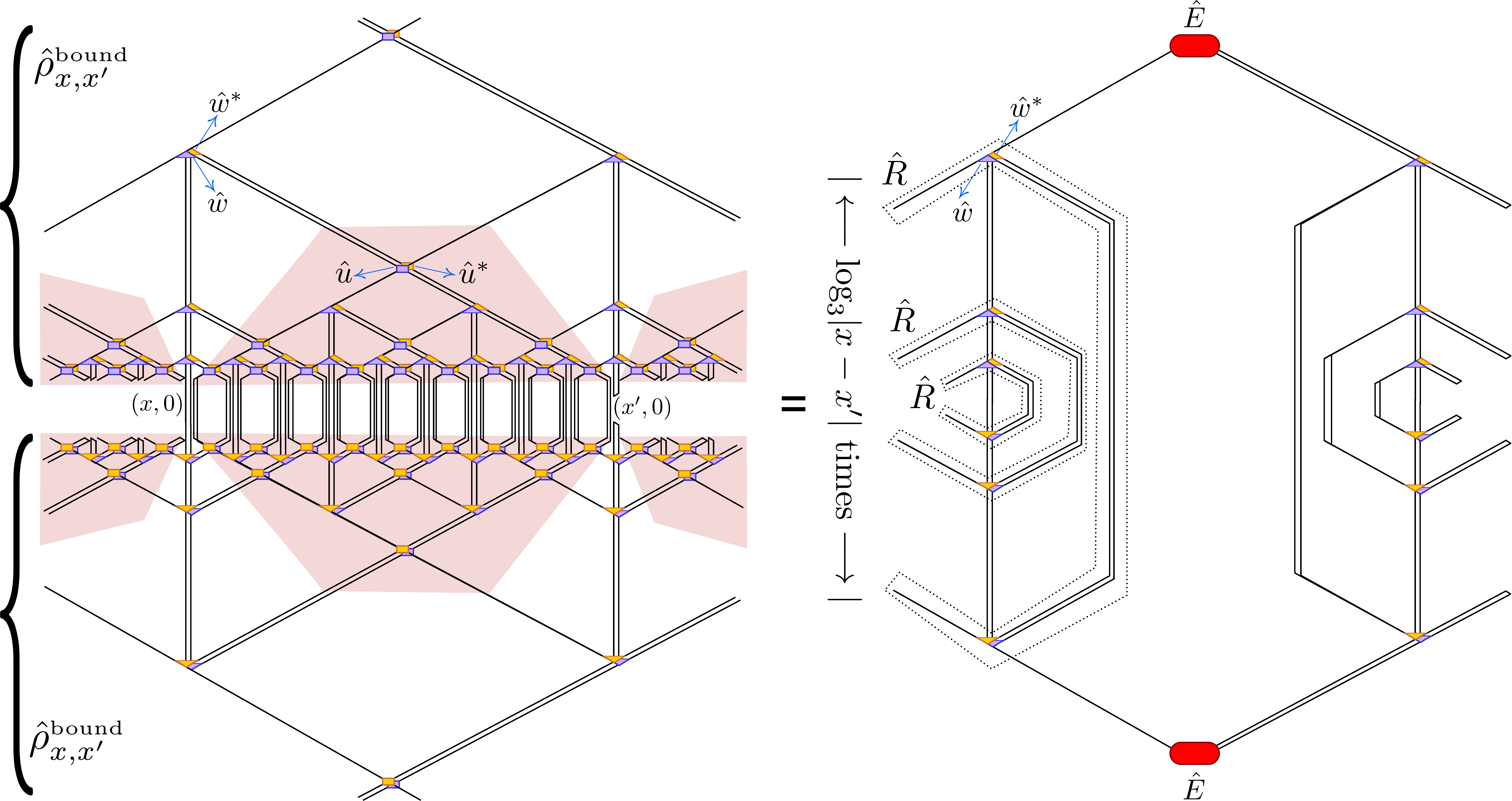}
\caption{\label{fig:entropy} (Color online) Tensor network contraction that equates to $[\mbox{Tr}(\hat{\rho}^{\tiny \mbox{bound}}_{x,x'})]^2$, where $\hat{\rho}^{\tiny \mbox{bound}}_{x,x'}$ is the boundary reduced density matrix of a block of $|x-x'|$ sites. The contraction consists of gluing together two copies of the tensor network representation of $\hat{\rho}^{\tiny \mbox{bound}}_{x,x'}$ along the open indices. Tensors located inside the red highlighted regions are pairwise mutiplied with their Hermitian adjoints and cancel out. The resulting contraction (shown on the right) consists of log $|x-x'|$ powers of a transfer superoperator $\hat{R}$ (obtained by contracting two copies of $\hat{w}$ and two copies of $\hat{w}^*$ that are enclosed within a dashed contour) and the same environment tensor $\hat{E}$ that appears in \fref{fig:corr}.}
\end{figure*}

\begin{figure*}
  \includegraphics[width=18cm]{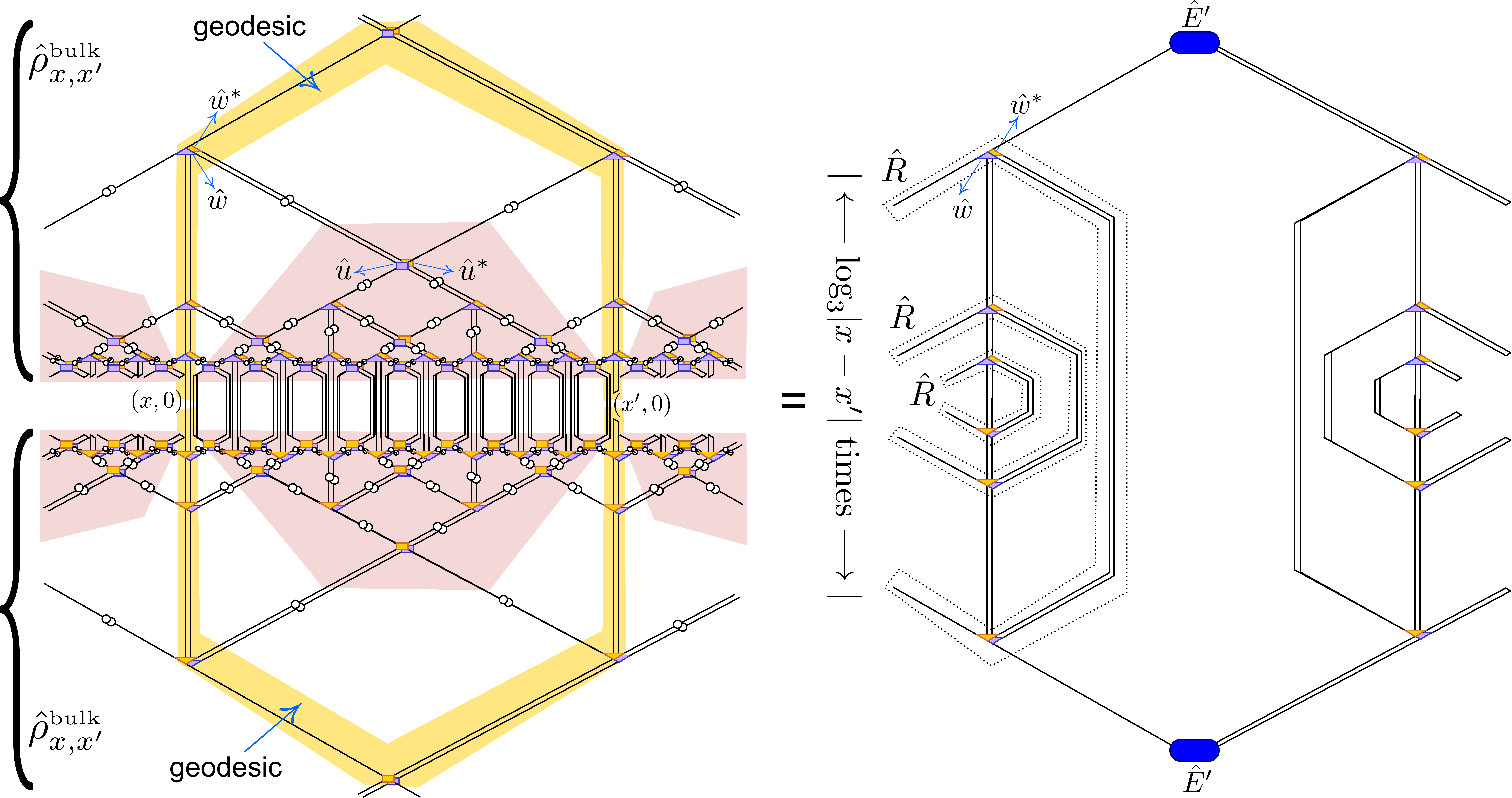}
\caption{\label{fig:bulkentropy}  (Color online) Tensor network contraction that equates to $[\mbox{Tr}(\hat{\rho}^{\tiny \mbox{bulk}}_{x,x'})]^2$, where $\hat{\rho}^{\tiny \mbox{bulk}}_{x,x'}$ is the bulk reduced density matrix of the bulk region $\mathcal{R}^{\tiny \mbox{bulk}}$ obtained from the tensor network representation of the projected bulk state $\myomega$, \eref{eq:omegaState}. The latter is obtained by applying the string projector $\hat{K}_{x,x'}$ on the bulk state $\myphi$. Operator  $\hat{K}_{x,x'}$ acts to remove the copy tensors located in the yellow highlighted regions. The contraction consists of gluing together two copies of the tensor network representation of $\hat{\rho}^{\tiny \mbox{bulk}}_{x,x'}$ along the open indices. Tensors located inside the red highlighted regions are pairwise mutiplied with their Hermitian adjoints and cancel out. The resulting contraction (shown on the right) consists of log $|x-x'|$ powers of the transfer superoperator $\hat{R}$ that appears in \fref{fig:entropy} but a different environment tensor $\hat{E}'$ (which also appears in \fref{fig:bulkcorr}).}
\end{figure*}

\subsection{3-point correlators}
The 3-point boundary correlator $\xbound{\hat{o}_\alpha(x) \hat{o}_\beta({x'}) \hat{o}_\gamma({x''})}$ is obtained by contracting a tensor network illustrated on the left hand side in \fref{fig:ope}. The contraction simplifies to the one depicted on the right hand side. For $\ell \equiv |x-x'| = |x'-x''| = |x''-x| = 3^{q}$, we obtain the closed expression
\begin{equation}\label{eq:corr3bound}
\begin{split}
\xbound{\hat{o}_\alpha(x) \hat{o}_\beta({x'}) \hat{o}_\gamma({x''})} = 
\frac{C_{\alpha\beta\gamma}}{\ell^{\Delta_\alpha + \Delta_\beta + \Delta_\gamma}},
\end{split}
\end{equation}
where $C_{\alpha\beta\gamma} = \mbox{tr}[\hat{F}(\hat{o}_\alpha \otimes \hat{o}_\beta \otimes \hat{o}_\gamma)]$.

Next, the corresponding bulk expectation value $\xbulk{\hat{T}_{x,x',x''} \hat{o}_\alpha(x,0) \hat{o}_\beta({x',0}) \hat{o}_\gamma({x'',0})}$ is obtained by contracting the tensor network depicted in \fref{fig:bulkope}. We obtain the closed expression
\begin{equation}\label{eq:corr3bulk}
\begin{split}
\xbulk{\hat{T}_{x,x',x''} \hat{o}_\alpha(x,0) \hat{o}_\beta({x',0}) \hat{o}_\gamma({x'',0})} =
\frac{C'_{\alpha\beta\gamma}}{\ell^{\Delta_\alpha + \Delta_\beta + \Delta_\gamma}},
\end{split}
\end{equation}
where $C'_{\alpha\beta\gamma} = \mbox{tr}[\hat{F}'(\hat{o}_\alpha \otimes \hat{o}_\beta \otimes \hat{o}_\gamma)]$.
Comparing \eref{eq:corr3bound} and \eref{eq:corr3bulk} we have
\begin{equation}
\begin{split}
\xbound{\hat{o}_\alpha(x) \hat{o}_\beta({x'}) \hat{o}_\gamma({x''})} =  g(\hat{u},\hat{w},\hat{o}_\alpha, \hat{o}_\beta, \hat{o}_\gamma) \times &\\
\xbulk{\hat{T}_{x,x',x''} \hat{o}_\alpha(x,0) \hat{o}_\beta({x',0}) \hat{o}_\gamma({x'',0})},&
\end{split}
\end{equation}
where $g(\hat{u},\hat{w},\hat{o}_\alpha, \hat{o}_\beta, \hat{o}_\gamma) = C_{\alpha\beta\gamma}/C'_{\alpha\beta\gamma}$, which is \eref{eq:OPE}.

\subsection{Reyni entanglement entropy}\label{ssec:entropyProof}

Figure \ref{fig:entropy} illustrates a tensor network contraction that equates to the second Reyni entanglement entropy $S^{(2)}(\mypsi,\mathcal{R}^{\tiny \mbox{bound}})$ of a block of boundary sites located at $x,x+1,\ldots,x'$.
In the limit of large $|x-x'|$ we obtain
\begin{equation}
S^{(2)}(\mypsi,\mathcal{R}^{\tiny \mbox{bound}}) \approx \mbox{log }\eta^{2\mbox{\tiny log }|x-x'|} + [\mbox{Tr}(\hat{\eta}\hat{E}\hat{\eta})]^2,
\end{equation}
where $\eta$ denotes the largest eigenvalue of transfer superoperator $\hat{R}$ obtained by contracting two copies of $\hat{w}$ and two copies of $\hat{w}^*$ that are enclosed within a dashed contour in \fref{fig:entropy}, $\hat{\eta}$ denotes the corresponding eigenoperator and $\hat{E}$ is the same environment tensor that appears in \eref{eq:corr2bound}. Here $\eta$ is related to the central charge of the CFT as $2\mbox{log }\eta=c/6$ \cite{ReyniCFT}.
Define the projected bulk state
\begin{equation} \label{eq:omegaState}
\myomega \equiv \hat{K}_{x,x'}\myphi.
\end{equation}
Denote by $\mathcal{R}^{\tiny \mbox{bulk}}$ the bulk region composed of sites enclosed between the geodesic extending between sites at $(x,0)$ and $(x',0)$ and the boundary, and including the bulk sites located at $(x+1,0),(x+2,0),\ldots,(x'-1,0)$.
The second Reyni entanglement entropy $S^{(2)}(\myomega,\mathcal{R}^{\tiny \mbox{bulk}})$ is obtained by contracting the tensor network illustrated in \fref{fig:bulkentropy}. In the limit of large $|x-x'|$, we obtain
\begin{equation}\label{eq:bulkreyni}
S^{(2)}(\hat{K}_{x,x'}\myphi,\mathcal{R}^{\tiny \mbox{bulk}})  \approx \mbox{log }\eta^{2\mbox{\tiny log }|x-x'|} + [\mbox{Tr}(\hat{\eta}\hat{E}'\hat{\eta})]^2,
\end{equation}
where $\hat{E}'$ is the environment tensor that appears in \eref{eq:corr2bulk}.
Comparing \eref{eq:reyni} and \eref{eq:bulkreyni} we have
\begin{equation}
S^{(2)}(\mypsi,\mathcal{R}^{\tiny \mbox{bound}}) = S^{(2)}(\myomega,\mathcal{R}^{\tiny \mbox{bulk}}) +  h(\hat{u},\hat{w}),
\end{equation}
where $h(\hat{u},\hat{w}) = [\mbox{Tr}(\hat{\eta}\hat{E}\hat{\eta})]^2 - [\mbox{Tr}(\hat{\eta}\hat{E}'\hat{\eta})]^2$, which is \eref{eq:reyni}.
This derivation readily generalizes to higher Reyni entropies, $\alpha > 2$. We obtain
\begin{equation}
S^{(\alpha)}(\mypsi,\mathcal{R}^{\tiny \mbox{bound}}) = S^{(\alpha)}(\myomega,\mathcal{R}^{\tiny \mbox{bulk}}) +  h^{(\alpha)}(\hat{u},\hat{w}),
\end{equation}
where $h^{(\alpha)}(\hat{u},\hat{w}) = [\mbox{Tr}(\hat{\eta}\hat{E}\hat{\eta})]^\alpha - [\mbox{Tr}(\hat{\eta}\hat{E}'\hat{\eta})]^\alpha$.


\begin{thebibliography}{74}
\bibitem{BulkCount} A finite MERA tensor network for a lattice of $N=2.3^{p}$ (boundary) sites has $N_B = 3^p + 2.3^{p-1} -3$ bond indices (which correspond to the number of bulk sites). Thus, if we view the infinite MERA as the large lattice size limit of a finite MERA, then the number of sites in $\mathcal{M}$ scale linearly with the number of sites in $\mathcal{L}$.
\end{thebibliography}
\end{document}